\documentclass[%
 aip,
 amsmath,amssymb,
reprint,%
]{revtex4-1}
\usepackage[utf8]{inputenc}
\usepackage[T1]{fontenc}
\usepackage{mathptmx}
\usepackage{etoolbox}
\usepackage{lipsum}
\usepackage{graphicx}
\usepackage{amssymb}
\usepackage{url}
\usepackage{soul}
\usepackage{xcolor}
\usepackage[labelformat=simple]{subcaption}

\usepackage{float}
\usepackage{calligra}
\usepackage{calrsfs}
\usepackage{xcolor}
\usepackage{enumitem}
\usepackage{graphicx}
\usepackage{dcolumn}
\usepackage{bm}
\usepackage{tabularx,ragged2e,booktabs,caption}
\usepackage{xcolor}
\usepackage{comment}
\usepackage[left=2cm,right=2cm,top=2cm,bottom=2cm]{geometry}

\definecolor{LWcolor}{RGB}{0,0,255} 

\newcommand{\mycomment}[1]{}
\usepackage{icomma}

\setcitestyle{numbers,square}

\definecolor{HGcolor}{RGB}{0,0,0} 
\newcommand{\hg}[1]{{\color{HGcolor}{#1}}}

\definecolor{GScolor}{RGB}{0,0,0} 

\makeatletter
\def\@email#1#2{%
 \endgroup
 \patchcmd{\titleblock@produce}
  {\frontmatter@RRAPformat}
  {\frontmatter@RRAPformat{\produce@RRAP{*#1\href{mailto:#2}{#2}}}\frontmatter@RRAPformat}
  {}{}
}%
\makeatother
\begin{document}
\title{Large Eddy Simulations of Flow over Additively Manufactured Surfaces: Impact of Roughness and Skewness on Turbulent Heat Transfer}

\author{Himani Garg}
 \affiliation{Lund University$,$ Department of Energy Sciences$,$ SE-22100 Lund$,$ Sweden}
\author{Guillaume Sahut}
\thanks{Currently employed by Cambridge Flow Solutions Ltd., United Kingdom.}
\affiliation{Lund University$,$ Department of Energy Sciences$,$ SE-22100 Lund$,$ Sweden}
\author{Erika Tuneskog}
  \affiliation{Chalmers University of Technology $,$ Department of Industrial and Materials Science $,$ SE-41296 Gothenburg$,$ Sweden}
\author{Karl-Johan Nogenmyr}
  \affiliation{Siemens Energy AB$ , $ SE-61231 Finspång$,$ Sweden}
\author{Christer Fureby}
\email{himani.garg@energy.lth.se; himani.garg65@gmail.com}
 \affiliation{Lund University$,$ Department of Energy Sciences$,$ SE-22100 Lund$,$ Sweden}

\begin{abstract}
\section*{Abstract}
\hg{Additive manufacturing creates surfaces with random roughness, impacting heat transfer and pressure loss differently than traditional sand-grain roughness. Further research is needed to understand these effects. We conducted high-fidelity heat transfer simulations over three-dimensional additive manufactured surfaces with varying roughness heights and skewness. Based on an additive manufactured Inconel 939 sample from Siemens Energy AB, we created six surfaces with different normalized roughness heights, $R_a/D = 0.001, 0.006, 0.012, 0.015, 0.020,$ and $0.028$, and a fixed skewness, $s_k=0.424$. Each surface was also flipped to obtain negatively skewed counterparts ($s_k=-0.424$). Simulations were conducted at a constant Reynolds number of 8000 and with temperature treated as a passive scalar (Prandtl number of 0.71). We analyzed temperature, velocity profiles and heat fluxes to understand the impact of roughness height and skewness on heat and momentum transfer. The inner-scaled mean temperature profiles are of larger magnitude than the mean velocity profiles both inside and outside the roughness layer. This means the temperature wall roughness function, $\Delta \Theta^+,$ differs from the momentum wall roughness function, $\Delta U^+$. Surfaces with positive and negative skewness yielded different estimates of equivalent sand-grain roughness for the same $R_a/D$ values, suggesting a strong influence of slope and skewness on the relationship between roughness function and equivalent sand-grain roughness. Analysis of the heat and momentum transfer mechanisms indicated an increased effective Prandtl number within the rough surface in which the momentum diffusivity is larger than the corresponding thermal diffusivity due to the combined effects of turbulence and dispersion. Results consistently indicated improved heat transfer with increasing roughness height and positively skewed surfaces performing better beyond a certain roughness threshold than negatively skewed ones.}
\end{abstract}  

\maketitle
\section{Introduction}\label{sec1}
Additive Manufacturing (AM) is being pursued for the demanding task of manufacturing gas turbine hot gas path components. The challenge comes from precisely maintaining the component at the highest temperature the material can tolerate given the load they are exposed to, such that the benefits in increased thermal efficiency are harvested. To develop such components, precise control over heat transfer and pressure losses in the cooling system is crucial. To this end, the studies detailed in this and previous papers have been conducted~\cite{bunker2017evolution,adair2019additive}.

AM surfaces, particularly Powder Bed Fusion Laser Beam (PBF-LB), exhibit high average surface roughness, typically falling within the range of $R_a=5-25\mu$m. The high surface roughness of PBF-LB surfaces stems from the powder-based manufacturing process, where a high-power laser selectively melts and solidifies layers of powder material to fabricate components \cite{Chowdhury2022LaserPB}. The intricate interplay of melt pool dynamics, laser angle of incident, print orientation, and other print process parameters contributes significantly to the formation of surface roughness during the AM process \cite{Ngo2018AdditiveM}. The topography of PBF-LB surfaces is characterized by the presence of partly sintered powder particles or larger spatter particles from the melt pool on the surface, accompanied by an underlying waviness originating from the melt pools. These particles manifest as sharp peaks in surface roughness measurements, often resulting in positively skewed surfaces. In addition, high peak density or variations in melt pool characteristics can result in valley-dominated surfaces.

Flow and heat transfer in channels have been studied for over a century. For rough and smooth walls, reliable correlations are available for pressure losses~\cite{colebrook1939correspondence,schlichting1937experimental,nikuradse1950laws,clauser1954turbulent}. For heat transfer, the situation is more complex. Heat transfer in internal air-cooled systems is governed by Newton's law of cooling, $Q=A \alpha (T_w - T_{cool})$. 
Typically, in a gas turbine context, the temperature difference between the cooling flow ($T_{cool}$) and the wall temperature ($T_w$) is given. However, the cooling engineer can optimize the design by adjusting the exposed surface area ($A$) and the heat transfer coefficient ($\alpha$). The heat transfer coefficient is favored by "thin flow passages," characterized by the hydraulic diameter, $D_h$, which can be seen directly in the definition of the Nusselt number, $Nu={\alpha D_h/\lambda}$, where $\lambda$ is the thermal conductivity of the fluid. Maximizing the surface area while minimizing the hydraulic diameter can be achieved with a cooling design that relies on in-wall cooling (or mini-) channels. 

Smooth channels are thoroughly explored, and well-established correlations are found (predominately Dittus-Bölter and Gnielinski \cite{dittus1930heat,gnielinski1975neue}). 
Regarding the heat transfer dynamics over rough surfaces, various empirical studies by Nunner~\cite{nunner1958heat}, Dipprey \& Sabersky~\cite{DIPPREY1963329}, and Kays \& Crawford~\cite{kays1980convective} have investigated predictive correlations for the Stanton number ($St$). Nunner~\cite{nunner1958heat} conducted pioneering experiments on a surface roughened with two-dimensional transverse ribs, providing an empirical expression for $St$ in terms of the Reynolds number, Prandtl number, and the ratio of rough to smooth skin friction coefficients. Additionally, Dipprey \& Sabersky~\cite{DIPPREY1963329} proposed a semi-analytical expression for $St$ based on the law of the wall similarity, incorporating the Prandtl number, skin friction coefficient, and inner-scaled equivalent sand-grain roughness, $k_s$. They addressed the challenge of incorporating the effects of $k_s$ on $St$ by leveraging experimental data from a pipe with sand-grain roughness. Subsequent studies have refined and modified these expressions based on experimental and numerical investigations~\cite{kays2005convective,bons2002st}.
Furthermore, due to challenges in obtaining accurate temperature fields within the roughness sublayer, the underlying physics of how wall roughness influences heat transfer remains ambiguous.

The effect of surface roughness has also been investigated numerically~\cite{li2007experimental,pelevic2016heat,lu2020effects,ansari2020influence,croce2007three,xiong2010investigation,kadivar2022cfd,kadivar2023comparison,garg2023largeribs}, with various methods. Several studies have proposed roughness models to account for roughness in laminar flows. Advancements in computer technology have facilitated Direct Numerical Simulations (DNS) for studying turbulent heat transfer over resolved rough surfaces~\cite{forooghi2018direct,peeters2019turbulent,macdonald2019heat,kuwata2021direct,garg2024heat}. MacDonald \textit{et al.}~\cite{macdonald2019heat} conducted DNS specifically on turbulent heat transfer over sinusoidally rough surfaces, indicating that the correlation function proposed by Dipprey \& Sabersky~\cite{DIPPREY1963329} accurately captured the influence of equivalent sand-grain roughness on $St$. Analysis of instantaneous temperature fields revealed that dissimilarities between heat and momentum transfer were primarily due to pressure drag effects on the rough wall, enhancing momentum transfer but not heat transfer. Peeters \& Sandham~\cite{peeters2019turbulent} explored DNS for grit-blasted surfaces, affirming the efficacy of Dipprey and Sabersky's correlation~\cite{DIPPREY1963329} in predicting $St$ for such surfaces. Notably, dissimilarities between heat and momentum transfers were evident in the recirculation zone behind roughness elements, where effective Prandtl numbers increased rapidly within the rough surface.  Recent DNS studies~\cite{macdonald2019heat,kuwata2021direct} have further investigated the scaling behavior of the temperature roughness function, $\Delta \Theta^+$, analogous to wall roughness function, $\Delta U^+$, and observed notable deviations from the established $\Delta U^+$ correlation, particularly at higher $k_s^+$ values.

In recent times, substantial efforts have been dedicated to investigating the relationship between topological roughness parameters and the consequent increase in friction factor ($c_f$). However, the surface roughness generated by AM techniques exhibits spatial non-uniformity, influenced by various factors such as profile curvature, layer thickness, laser power, sample orientation, metallic composition, and particle size. As a result, AM roughness differs considerably from regular, random, and artificial roughness. To date, convective heat transfer in AM-made mini-channels remains understudied, with limited experimental and numerical investigations \citep{stimpson2017scaling,snyder2020tailoring,mcclain2021flow,FAVERO2022106128,garg2024heat}. Snyder \textit{et al.} \cite{snyder2020tailoring} demonstrated successful tailoring of surface roughness through AM process parameters, enhancing the performance of a generic micro-channel cooling design. More recent studies by Garg \textit{et al.}~\cite{garg2023large,garg2024heat} have largely focused on systematically exploring the relationship between AM roughness characteristics and $k_s$ to reveal their impact on turbulent heat transfer using the similarity function, all-normal Reynolds stress, heat flux, and probability density functions by using wall resolved Large Eddy Simulations (LES). It has been explored how turbulence statistics is affected by the presence of the roughness elements, particularly "peaks" dominant surfaces, and the highly non-uniform heat transfer was shown to predominately take place at the "peaks" of the wall roughness. In this paper, we continue these studies and aim to show higher-order statistics of the roughness characteristics that affect heat transfer by directly comparing "peak" and "valleys" dominant surfaces. To the best of the authors' knowledge, this study represents the first instance of using LES to simulate turbulent heat transfer in "peak" and "valley" dominated AM rough wall pipe flows with grid-conforming three-dimensional roughness elements. 
\begin{figure}
\begin{subfigure}[h]{0.55\columnwidth}
\includegraphics[width=\linewidth]{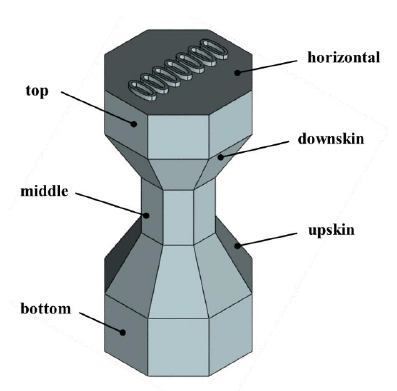}
\end{subfigure}
\hfill
\begin{subfigure}[h]{0.4\linewidth}
\includegraphics[width=\linewidth]{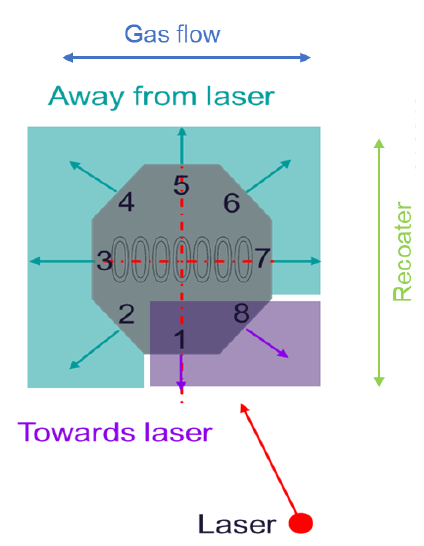}
\end{subfigure}%
\caption{Illustration of the test specimen. (a) Naming of different levels. (b) Test specimen in relation to gas flow, reacoater, and laser.}
\label{fig: test_obj}
\end{figure}
\section{Geometry creation}
\subsection{Experimental}
The test specimen used for surface roughness measurements was manufactured using PBF-LB on an EOS M400-4 system at Siemens Energy AB, Finspång. The powder feedstock was Inconel 939, with a powder particle size distribution ranging from 15 to 45 $\mu$m. Printing parameters were set according to Siemens Energy standard and the inert gas was argon. The build direction was set from bottom to top, as shown in Fig. \ref{fig: test_obj}. The specimen had a height of 30mm and a maximum diameter of 15mm, featuring a total of 41 surfaces with orientations at -60$^{\circ}$ (downskin), 0$^{\circ}$, 60$^{\circ}$ (upskin), and 90$^{\circ}$. Each orientation (except 0$^{\circ}$) comprised a minimum of 8 surfaces. Surface measurements were performed on each surface using both an optical high-resolution light microscope, the Leica DM6 M, with focus variation, and an SEM, the Jeol JSM-IT500. Post-processing, encompassing profile, and areal evaluation were carried out using the Leica Map DCM software (Release: 7.4.8964). 
\label{Sec: 3.1}
 \begin{table*}
\begin{center}
\caption{Summary of roughness parameters based on roughness height, $y$ (assuming mean line is at $y=0$), with continuous formulations, where $L_x$ is the surface length, and $p$, the probability density function.}
\begin{tabular}{l p{60mm} l l}
\hline 
Parameter & Description & Continuous formulation   \\ \hline \\[0.05cm]
$R_a$   & Arithmetic average height [m] & $\frac{1}{L_xL_z}\int_0^{L_x}\int_0^{L_z} \left|y(x,z)\right | dzdx$    \\[0.5cm]
$R_q$   & Root-mean-square roughness height [m]  & $\sqrt{\frac{1}{L_xL_z}\int_0^{L_x}\int_0^{L_z} \left \{ y(x,z)\right \}^2 dzdx}$        \\[0.5cm]
$R_z$   &  Maximum height between the highest peak and the deepest valley of the profile [m]  & $\left |\min y_i\right |+ \left|\max y_i \right|$ \\[0.5cm] 
$s_k$   & Skewness [-]  & $\frac{1}{R_q^3} \int_{-\infty}^{+\infty} y^3p(y)dy$         \\[0.5cm] 
$k_u$   & Kurtosis [-]& $\frac{1}{R_q^4} \int_{-\infty}^{+\infty}y^4p(y)dy$       \\[0.5cm]  \hline
\end{tabular}
\label{table: Roughness parameters intro}
\end{center}
\end{table*}

\begin{figure*}[!htb]
\begin{center}
\includegraphics[width=0.9\linewidth]{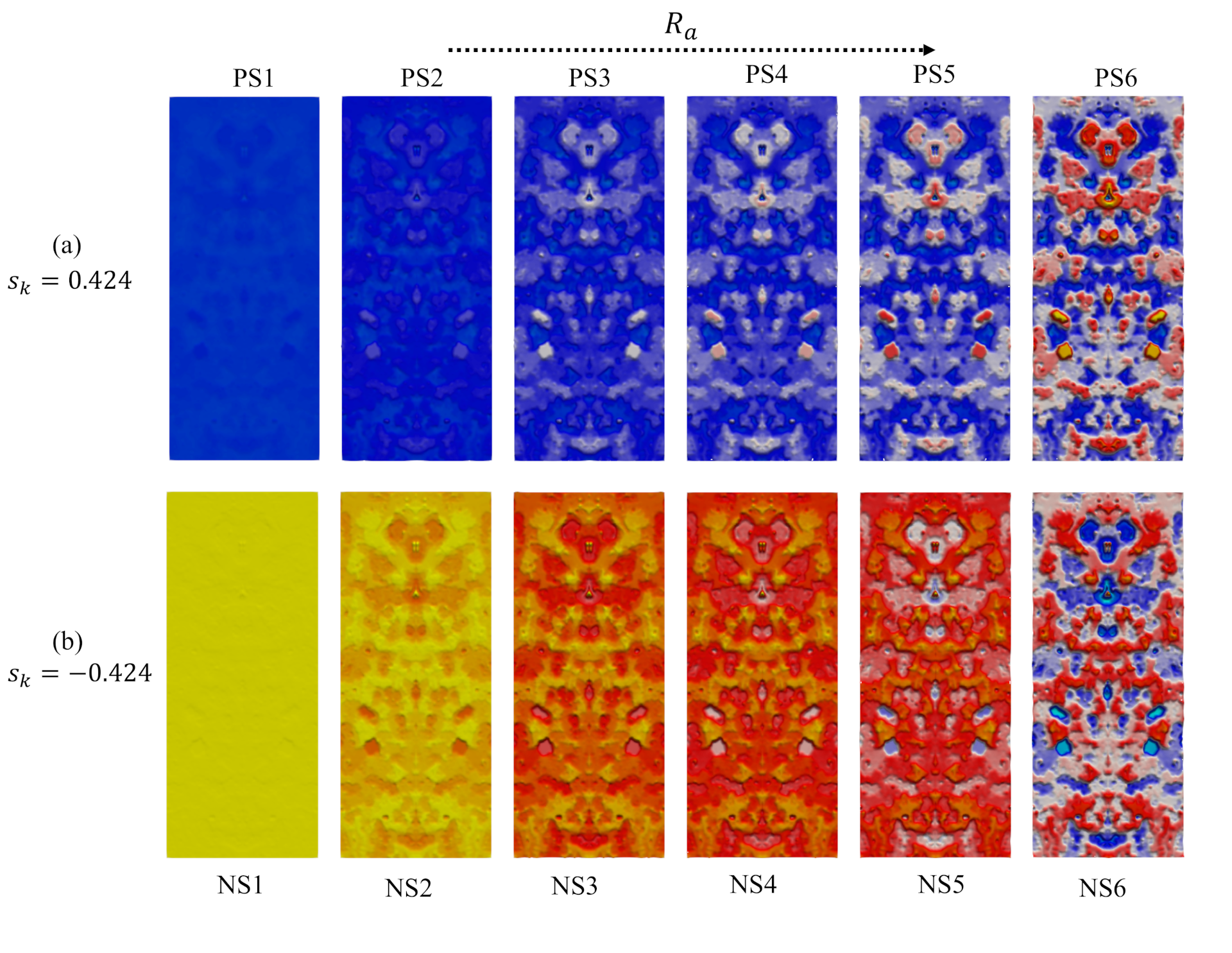}
\caption{Visualization of additive manufacturing rough surface height map, $y(x,z)$, extracted from the measurement of 3D printed microchannels at Siemens Energy AB. The surfaces in the top row have $s_k=0.424$; those in the bottom row have $s_k=-0.424$ value. From left to right, the sampled surface is the same, but the mean roughness height increases. 
}
\label{fig: Rough Surface Heigh Map}
\end{center}
\end{figure*}
\subsection{Rough surface characterization}
Our goal in this study is to directly isolate the effects of roughness height and skewness on turbulent heat transfer. To achieve this, we needed to tightly control the roughness characteristics of the surfaces we investigated 
by using a specific rectangular area from the downskin surface at position 5 (see Fig.~\ref{fig: test_obj}) as a base. We then resized this base area to create six samples with varying roughness heights, while keeping the underlying surface pattern constant. The original surface 
was positively skewed, i.e., dominated by peaks. To create samples with negative skewness, we simply flipped these original surfaces upside down. In total, twelve rough surfaces were generated based on the data provided by Siemens Energy AB, shown in Fig.~\ref{fig: Rough Surface Heigh Map}. Flipping the surfaces essentially transformed the peaks into valleys, all while maintaining consistent roughness parameters. Table~\ref{table: Roughness parameters intro} summarizes the key characteristics derived from the probability density function (PDF) of the surface height. These characteristics include the average roughness height, $R_a$, the maximum peak-to-valley height, $R_z$, the skewness factor, $s_k$, and the kurtosis factor, $k_u$, all calculated using the height function, $y(x,z)$, over sampling lengths $L_x$ and $L_z$ in the streamwise and spanwise directions, respectively. We denote surfaces with positive skewness (peak-dominated) as PS and those with negative skewness (valley-dominated) as NS for brevity.
 \begin{table}
\begin{center}
\caption{Statistical quantities for the twelve rough surfaces considered. The descriptions and analytical expressions are given in Table \ref{table: Roughness parameters intro}. Here $D$ is the characteristic diameter of the rough pipe (defined later).}
\begin{tabular}{p{10mm} p{15mm} p{15mm} p{15mm} c}
\hline \\[0.cm]
Case  & $R_a/D$ [-] & $R_z/D$ [-]  & $s_k$ [-] & $k_u$ [-]  \\[0.2cm]\hline
PS1    & 0.001 & 0.010  & 0.424   & 3.054  \\[0.2cm]
PS2    & 0.006 & 0.052  & 0.424   & 3.054  \\[0.2cm]
PS3    & 0.012 & 0.103  & 0.424   & 3.054  \\[0.2cm]
PS4    & 0.015 & 0.127  & 0.424   & 3.054  \\[0.2cm]
PS5    & 0.020 & 0.164  & 0.424   & 3.054   \\[0.2cm]
PS6    & 0.028 & 0.234  & 0.424   & 3.054   \\[0.2cm]
NS1    & 0.001 & 0.010  & -0.424  & 3.054   \\[0.2cm]
NS2    & 0.006 & 0.053  & -0.424  & 3.054   \\[0.2cm]
NS3    & 0.012 & 0.103  & -0.424  & 3.054   \\[0.2cm]
NS4    & 0.015 & 0.127  & -0.424  & 3.054  \\[0.2cm]
NS5    & 0.020 & 0.164  & -0.424  & 3.054  \\[0.2cm]
NS6    & 0.028 & 0.234  & -0.424  & 3.054   \\[0.cm]
\hline
\end{tabular}
\label{table: Roughness parameters values}
\end{center}
\end{table}
%
Figure~\ref{fig: Rough Surface Heigh Map}(a) displays surfaces with a positive value of $s_k$, while Fig.~\ref{fig: Rough Surface Heigh Map}(b) depicts surfaces with a negative value of $s_k$. Going from left to right, the normalized roughness height ($R_a/D$) ranges from 0.001 to 0.028. The roughness parameters, determined using the code developed by Garg et al. \cite{garg2023large,garg2024heat}, are detailed in Table \ref{table: Roughness parameters values}. Notably, the values of $k_u$ remain constant across all rough surfaces examined in this study.
\subsection{Geometries and meshes from experiment}
\label{sec: geometry creation}
\begin{figure}[b]
\begin{center}
%
\includegraphics[width=0.8\linewidth]{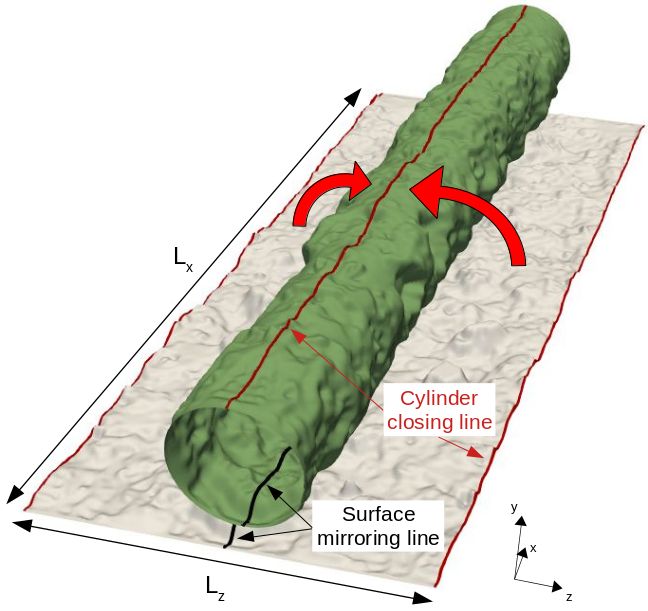}
   \caption{A planar rough surface mirrored with respect to the black line and then wrapped around the same direction to produce a rough pipe.}
   \label{fig:code_plane2cylinder}
   \end{center}
\end{figure}

A rectangular portion of the downskin surface at position 5, as shown in Fig. \ref{fig: test_obj}, was extracted and written in STL format. Figure \ref{fig:code_plane2cylinder} shows the transformation of this rough plane to a rough pipe. The planar surface is mirrored along the $x$-axis. This operation ensures that both sides of the surface parallel to this axis are identical. Then a rotation around the $x$-axis is applied to all points of the STL file. Along the closing line, all pairs of points merge perfectly thanks to the previous mirroring step, as shown in Fig. \ref{fig:code_plane2cylinder}.

Practitioners can easily bend a planar surface using commercial and open-source CAD software. However, merging the points along the closing line is quite time-consuming and error-prone. Indeed, this operation relies on a threshold distance to decide whether two points on the closing line should be merged. The determination of this threshold value for a given STL file is cumbersome.

To ease the process of creating rough cylinders, the authors have developed an open-source Fortran code available on GitHub \footnote{\url{https://github.com/CoffeeDynamics/STLRoughPipes}}. The code runs on Unix systems and takes as input parameters the STL file containing the planar rough surface and a tolerance factor used to merge the points on the closing line. Optionally, the user can specify a roughness factor to rescale the surface roughness. Several values of tolerance factors may need to be tested in a trial-and-error approach in order to find the most appropriate one. A tolerance factor of $0.002$ was found accurate to properly merge the points on the closing line for all the pipes considered in this work. The flexibility of the code is a major asset for the creation of rough pipes from AM rough surfaces. Moreover, the ability to rescale the surface roughness leads to straightforward parametric studies of the impact of the roughness height and skewness on turbulence and heat transfer. The code produces an STL file containing the rough pipe, as shown in Fig. \ref{fig:code_plane2cylinder}, and related statistical quantities, as detailed in Table \ref{table: Roughness parameters intro} (only the ones of interest for the present work). This code has been used in our previous studies~\cite{garg2023large,garg2024heat}.
\begin{figure}[!htb]
\begin{center}
      \includegraphics[width=1\linewidth]{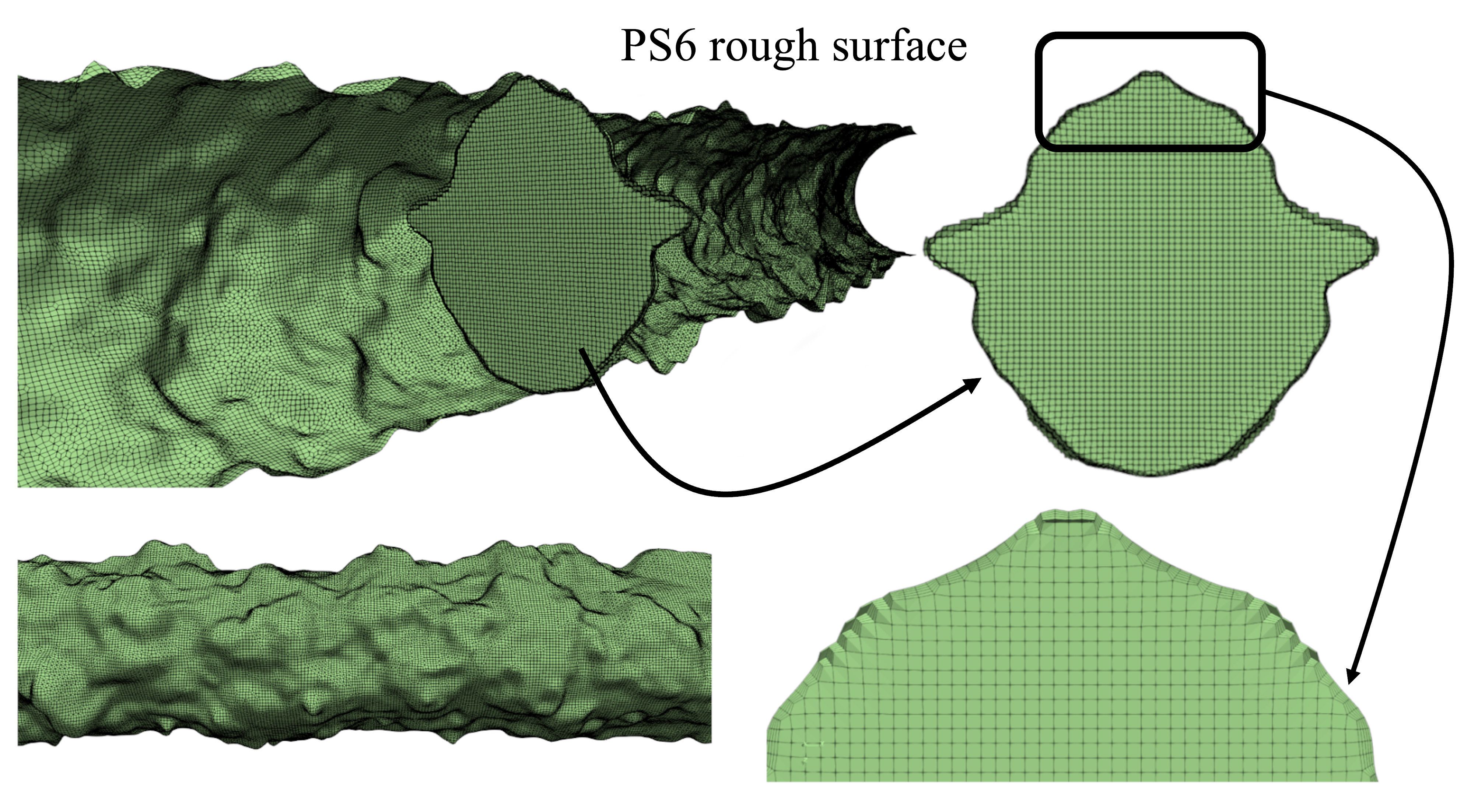}
   \caption{Illustration of the mesh for PS6 rough surface.}
   \label{fig:mesh for PS6}
   \end{center}
\end{figure}

We used snappyHexMesh, a meshing tool included in the OpenFOAM software (OF7), to create meshes for the volumes within the cylinders. As an example, Fig.~\ref{fig:mesh for PS6} shows the mesh generated for the PS6 rough surface. To precisely capture the surface roughness, we refined the mesh near the walls.  However, due to the intricate surface details and the complexity of cells close to the walls, obtaining exact values for non-dimensional wall distances like $x^+$, $r^+$, and $z^+$ proved challenging. Therefore, we present estimated average values for these quantities in our current meshes. Across all surfaces at a Reynolds number ($Re_b$) of 8000, the average values of $x^+$, $r^+$, and $z^+$ near the wall are all found to be below 2.35. In the channel center, $r^+$ values range from 3.36 to 10.90 for PS1 to PS6 and NS1 to NS6 roughness configurations. The maximum values of skewness, aspect ratio, and non-orthogonality for cells across PS1 to PS6 and NS1 to NS6 roughness are observed to be between 0.72 and 1.59, 10.37 and 13.41, and $51.56^\circ$ and $65.33^\circ$, respectively.
\section{Numerical methodology}
\label{sec:2}
\subsection{Governing equations}
This study involves conducting wall-resolved LES of thermal flow in pipes roughened with AM structures. The flow is assumed to be fully incompressible, Newtonian, and heated with a uniform wall heat flux $q_w$. The governing equations are the spatially filtered mass, momentum, and energy conservation equations, where the filtered quantities are denoted by overbars, with a subgrid-scale (SGS) model introduced to represent small-scale unresolved turbulence effects. The filtered LES equations are expressed as follows: 
\begin{equation}
    \frac{\partial \overline{u}_i}{\partial x_i}=0,
    \label{Eq: 1}
\end{equation}
\begin{equation}
    \frac{\partial \overline{u}_i}{\partial t} + \frac{\partial}{\partial x_j}(\overline{u}_i\overline{u}_j) = -\frac{1}{\rho}\frac{\partial\overline{p}}{\partial x_i}-\frac{\partial\tau_{ij}}{\partial x_j}+\frac{\partial}{\partial x_j}\left[\nu \left(\frac{\partial \overline{u}_i}{\partial x_j} \right) \right],
    \label{Eq: 2}
\end{equation}
\begin{equation}
\frac{\partial \overline{T} }{\partial t} + \frac{\partial \left(\overline{T}\bar{u}_i\right)}{\partial x_i}=-\frac{\partial q_i} {\partial x_i} + \frac{\partial}{\partial x_i}\left[\frac{\nu}{Pr}\frac{\partial \overline{T}}{\partial x_i}\right],
    \label{Eq: 3}
\end{equation}
where $\overline{{u}}_{i}$ is the filtered velocity, $\overline{p}$ the filtered pressure, $\nu$ the kinematic viscosity, and $\tau_{ij} = \overline{u_i u_j}-\overline{u_i}\,\overline{u_j}$ the subgrid-scale stress tensor, modeled using the Boussinesq hypothesis, $\tau_{ij}=-2\rho\nu_{sgs}\bar{S}_{ij}$, where $\bar{S}_{ij}$ is the resolved rate-of-strain tensor and $\nu_{sgs}$ is the subgrid viscosity. Additionally, $T$ denotes the temperature, $\alpha=\nu/Pr$, the thermal diffusivity, $Pr$, the Prandtl number, and $q_i = \overline{u_iT}-\overline{u_i}\overline{T}$, the subgrid heat flux vector. The subgrid viscosity is modeled using the Wall-Adapting Local Eddy-viscosity (WALE) model~\cite{nicoud1999subgrid}, given by:
\begin{equation}
\nu_{t} = (c_w\Delta)^2 \frac{(S^d_{ij}S^d_{ij})^{3/2}}{(\overline{S}_{ij}\overline{S}_{ij})^{5/2} + (S^d_{ij}S^d_{ij})^{5/4}},
\label{eq: WALE}
\end{equation}
where $c_w \simeq 0.325$ is a model parameter, $\Delta$ is the cell filter size, $S^d_{ij} = \frac{1}{2}(\overline{g}^2_{ij}+\overline{g}^2_{ji}) - \frac{1}{3}\delta_{ij}\overline{g}^2_{ij}$ is the traceless symmetric part of the square of the velocity gradient tensor, with $\overline{g}_{ij} = \partial \overline{u}_i / \partial x_j$ being the filtered velocity gradient. The subgrid heat flux is modeled with a gradient transport model~\cite{moin1991dynamic}, with $\alpha_{sgs}=\nu_{sgs}/Pr_t$, where $Pr_t$ is the turbulent $Pr$ number.

With a constant heat flux boundary condition, the fluid's bulk mean temperature linearly increases with axial distance, complicating the implementation of periodic boundary conditions for the temperature. To address this, we adopt the approach suggested by Kasagi \textit{et al.} \cite{kasagi1992}, introducing a transformed temperature $\Theta = T - T_w$ in the energy equation~\cite{garg2024heat}, where $T_w$ is the temperature at the wall. The advantage of this method lies in the periodicity of $\Theta$ in the streamwise direction \cite{kozuka2009,lluesma2018}. Under fully developed flow conditions, the energy equation is modified to incorporate an internal heat source as follows,
\begin{equation}
    \frac{\partial \overline{\Theta}}{\partial t}+\frac{\partial \left(\overline{\Theta} \overline{u}_i\right)}{\partial x_i}=-\frac{\partial q_i^*} {\partial x_i}+\frac{\partial}{\partial x_i}\left[\frac{\nu}{Pr}\frac{\partial \overline{\Theta}}{\partial x_i}\right] +\frac{q_w u_\tau}{\rho c_p \nu_w}\frac{ \overline{u}_x}{ \left< \overline{u}_x\right>},
    \label{Eq: 4}
\end{equation}
where $u_\tau$ is the friction velocity and $q_i^* = \overline{u_i\Theta}-\overline{u_i}\overline{\Theta}$, the transformed subgrid heat flux vector.
\subsection{Numerical methods}
In this study, OF7 was used to simulate turbulent pipe flow with AM rough walls. The LES equations were solved numerically using a second-order cell-centered discretization scheme for convective and diffusive fluxes within the finite volume method. Time stepping involved the implicit Adams-Bashforth method~\cite{butcher2016numerical}, with a maximum Courant-Friedrichs-Lewy (CFL) number of 0.5 to ensure numerical stability. The Pressure Implicit Splitting of Operators (PISO) algorithm \cite{weller1998tensorial} was used to achieve pressure-velocity coupling, with three corrector steps to minimize discretization errors. The resulting pressure equation was solved using the Generalized Geometric-Algebraic Multigrid (GAMG) method \cite{GAMG} with a Diagonal Incomplete Cholesky (DIC) smoother \cite{van1996matrix}. For velocity solutions, the Preconditioned BiConjugate Gradient (PBiCG) solver \cite{press2007numerical} was employed along with a Diagonal Incomplete Lower-Upper (DILU) preconditioning method \cite{van1996matrix}. Additionally, to address non-orthogonality in our roughness-conforming meshes, two inner loop correctors were implemented, considering that the maximum non-orthogonality in all meshes remained below $70^\circ$. 
\subsection{Initial and boundary conditions, dimensionless numbers}
The simulations assume fully developed, turbulent, incompressible, and Newtonian flow with constant properties ($\rho=1$ kg m$^{-3}$ and $c_p=1005$ J kg$^{-1}$K$^{-1}$). The effects of gravitational acceleration forces are neglected. The value of $Pr$ was set to 0.71 and $Pr_t$ was set to 0.85. The simulations explored the influence of two key factors: roughness height and surface skewness. We simulated six different normalized roughness heights ($R_a/D$ = 0.001 to 0.028) on surfaces with two distinct skewness values ($s_k = -0.424$ and $0.424$), where $D$ is the characteristic diameter of the rough pipe defined as $D = S/\left( \pi L_x\right)$ with $S$ being the total rough surface area of the pipe and $L_x$ the pipe length. The $Re_b$ was kept constant at $8000$ throughout all simulations. Here, $Re_b$ is defined as $Re_b = U_bD/\nu$, where $U_b$ is the bulk velocity of the fluid.
For easy reference,  we labeled the surfaces with positive skewness ($s_k = 0.424$) as PS followed by a number indicating the roughness height (e.g., PS1 for $R_a/D = 0.001$). Similarly, surfaces with negative skewness ($s_k = -0.424$) were labeled NS followed by a number (e.g., NS6 for $R_a/D = 0.028$).

In analyzing wall-bounded flows, the friction velocity scale, $u_\tau$, is determined \textit{a posteriori} using the pressure gradient, $\Delta p/L_x$, calculated as $u_\tau =\sqrt{0.25D\Delta p/(\rho L_x)}$. The friction temperature, $\Theta_\tau$, is defined as $\Theta_\tau = {q_w}/{\rho c_pu_\tau}$. A key parameter, the turbulent friction $Re$ number, $Re_\tau$, is defined as $Re_\tau =u_\tau D/(2\nu)$. In this study, $Re_\tau$ ranges from 260 to 568 for NS1 to NS6 and 260 to 602 for PS1 to PS6. Additionally, the roughness Reynolds number, $k_s^+= {k_s u_\tau}/{\nu}$, is considered, where $k_s$ is estimated based on the roughness function, $\Delta U^+$~\cite{garg2023large,garg2024heat}. This yields a range of $k_s^+$ between 1 and 518 for negatively skewed surfaces and between 1 and 536 for positively skewed surfaces. Notably, in flows over roughened walls, $u_\tau$ and the skin friction coefficient, $c_f$, reflect the total wall drag, encompassing pressure and viscous drag instead of solely the skin-friction drag.

The computational domain length, $L_x$, is set to eight times the pipe diameter, i.e., $L_x=8D$, to mitigate periodicity effects in all considered statistics~\cite{lozano2014effect,lluesma2018influence,garg2023large,garg2024heat,garg2024large}. The axial, radial, and azimuthal directions are denoted by $x$, $r$, and $\theta$, respectively, with corresponding velocity components $u_x$, $u_r$, and $u_\theta$. The effective wall-normal distance is denoted by $r = r^0-d$, where $r^0$ is the average pipe radius and $d$ is the zero-plane displacement~\cite{garg2023large,garg2024heat}. Periodic boundary conditions are applied in the streamwise direction, while no-slip boundary conditions are imposed at the wall for all velocity components. The zero-Neumann boundary condition is used for pressure, with an additional forcing term incorporated to address the pressure gradient effect within the context of periodic boundary conditions. The boundary condition for the transformed temperature is simply $\Theta = 0$ at $r=0$ and $r=D$.

Statistical representation involves angular brackets, $\left<...\right>$, for time-averaged quantities and primes, $\left(...\right)'$, for fluctuating quantities. The reported data correspond to a dataset extending to $100t_{\text{ftt}}$, where $t_{\text{ftt}} = L_x/U_b$ is the flow-through time.
\section{\label{sec:3}Results and discussions}  
\subsection{Instantaneous flow}
\begin{figure*}[!ht]
\begin{center}
\subfloat[]{
\includegraphics[width=0.8\linewidth]{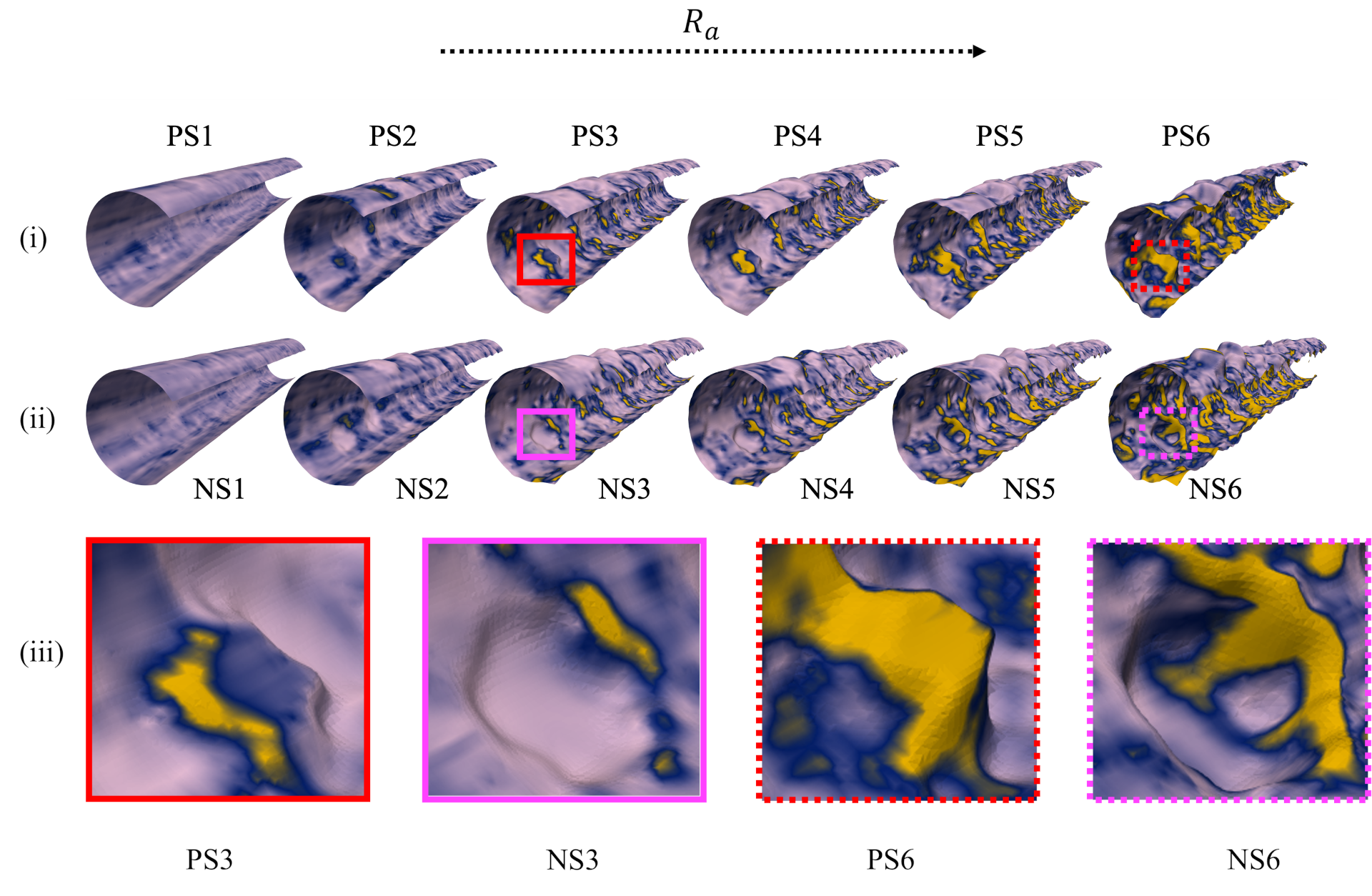}}

\subfloat[]{
\includegraphics[width=0.8\linewidth]{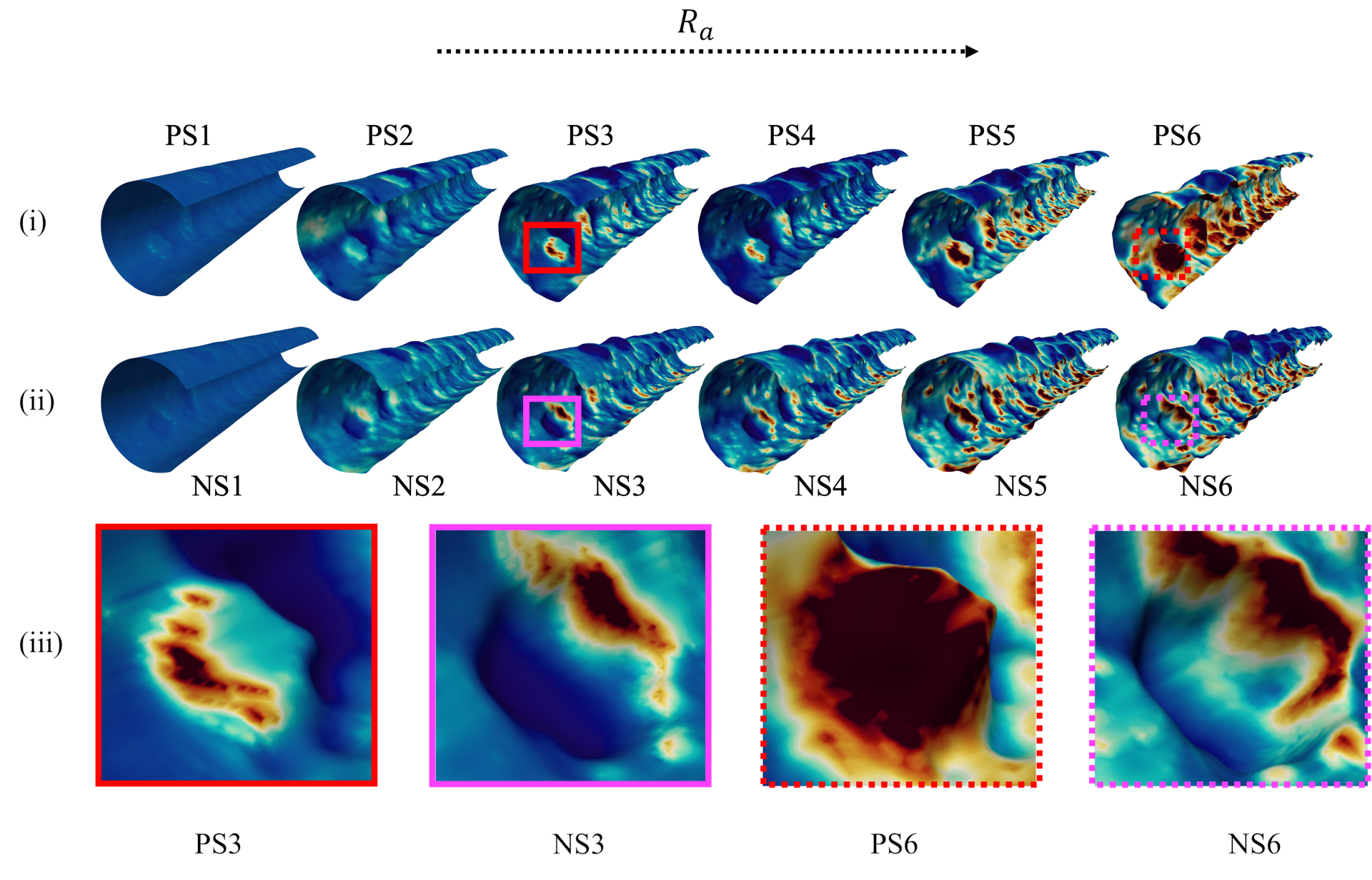}}
\caption{(a) Skin friction factor ($c_f$, colormap varies from pink (minimum) to yellow (maximum)) and (b) heat transfer enhancement factor ($Nu Pr^{-1/3}$, colormap varies from blue (minimum) to red (maximum)), as a function of mean roughness height, $R_a$, for (i) positively and (ii) negatively skewed rough surfaces. (iii) Zooms are also included for clarity.}
\label{fig: cf and Nu visualizations}
\end{center}
\end{figure*}
In the context of heat transfer over rough surfaces, the combined effect of increasing surface roughness and skewness can result in a synergistic enhancement of heat transfer. From a statistical sense, the surfaces with positive skewness can be considered peak-dominated surfaces, and the ones with negative skewness as valley-dominated surfaces.  To explore the influence of peaks and valleys alongside their height impact on the friction factor $c_f=\Delta p D/(2 L_x \rho U_b^2)$ and Nusselt-Prandtl ratios, firstly, the instantaneous flow visualizations of $c_f$ (absolute values) and $NuPr^{-1/3}$ are shown in Fig.~\ref{fig: cf and Nu visualizations}.

In Fig.~\ref{fig: cf and Nu visualizations}(a) and \ref{fig: cf and Nu visualizations}(b), moving from left to right, the $s_k$ values are fixed, but roughness height increases. Panels (i) and (ii) show the results for positively and negatively skewed roughness, respectively. For small values of $R_a/D$, the distribution of $c_f$ resembles the one obtained for smooth pipe flows, and the impact of $s_k$ is negligible. With increasing $R_a/D$ values, all surfaces show elevated $c_f$. Higher $c_f$ values at peaks indicate that the roughness elements are inducing more substantial energy losses in the flow. These energy losses are associated with the work done by the fluid to overcome the increased resistance caused by the rough surface. In geometries transitioning from peaks to cavities at a fixed $R_a/D$ ratio, changes in the distribution of $c_f$ along the surface are observed (see zooms included in panel (iii)). In peak geometries, such as protrusions, higher $c_f$ values are typically found at the front head of the peak due to flow separation and the formation of recirculation zones. However, when the geometry is rotated to form a cavity, the flow behavior alters, leading to flow reattachment downstream of the cavity. Consequently, the downstream side of the cavity experiences increased turbulence levels and higher momentum transfer near the surface, resulting in higher $c_f$ values on the downstream side of the cavity. These variations in flow behavior and boundary layer characteristics contribute to the observed changes in the distribution of $c_f$ as well as in heat transfer along the surface of the geometry, see Fig.~\ref{fig: cf and Nu visualizations}(b). The heat transfer enhancement factor exhibited a pronounced upward trend as the surface roughness increased. The presence of roughness elements altered the near-wall flow patterns, leading to increased turbulence and improved heat transfer rates. Furthermore, in peak geometries where $c_f$ values are higher at the front head of the peak, the enhanced turbulence near the surface promotes better mixing of the fluid, resulting in higher convective heat transfer coefficients. Consequently, $Nu Pr^{-1/3}$ tends to be relatively higher on the upstream side of the peak compared to the downstream side. 
\subsection{Mean velocity and temperature}
\begin{figure}[!ht]
\begin{center}
\subfloat[]{
\includegraphics[width=0.95\linewidth]{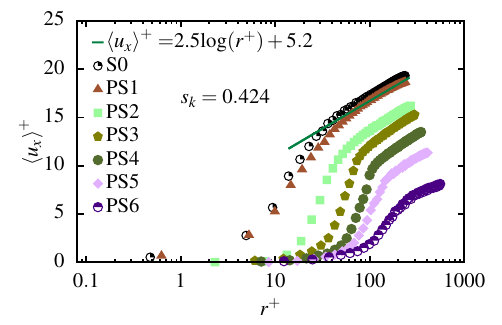}}

\subfloat[]{
\includegraphics[width=0.95\linewidth]{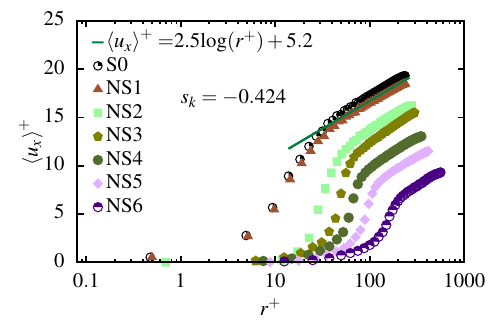}}
\caption{Mean velocity profile for (a) positively skewed and (b) negatively skewed surfaces in the inner wall units.}
\label{fig: Mean flow profile}
\end{center}
\end{figure}

Figure~\ref{fig: Mean flow profile} presents inner-scaled profiles of the streamwise mean velocity, $\left< u_x^+\right>$, for positively and negatively skewed surfaces. The non-dimensional wall distance, $r^+$, is measured from the plane at which the total drag acts. This involves shifting the velocity profiles for rough surfaces by removing the negative velocities, commonly referred to as zero-plane displacement~\cite{raupach1994simplified,garg2023large,garg2024heat}. Additionally, the plot includes numerical results of mean streamwise velocity profiles for LES of smooth pipe flow at the same bulk $Re$ number ($Re_b = 8000$) for comparison and standard log-layer profiles for smooth walls (solid green line). In Figs.~\ref{fig: Mean flow profile}(a) and \ref{fig: Mean flow profile}(b), it is observed that increasing $k_s^+$ results in a downward shift of the logarithmic profile of $\left< u_x^+\right>$, indicating enhanced momentum transfer by the wall roughness. For small values of $R_a/D$, i.e., in the hydraulically smooth regime as $k_s^+ \approx 1$, the PS1 and NS1 profiles are very close to the one for smooth surfaces, S0, as expected. The impact of $s_k$ is little for PS2, PS3, NS2, and NS3 surfaces as the downward shift is found to be roughly the same. For the larger value of $R_a/D > 0.015$, the downward shift in the logarithmic profile is found to be larger for positively skewed surfaces compared to negatively skewed ones, signifying larger momentum transfer by the wall roughness in the former ones.

\begin{figure}[!ht]
\begin{center}
\subfloat[]{
\includegraphics[width=0.95\linewidth]{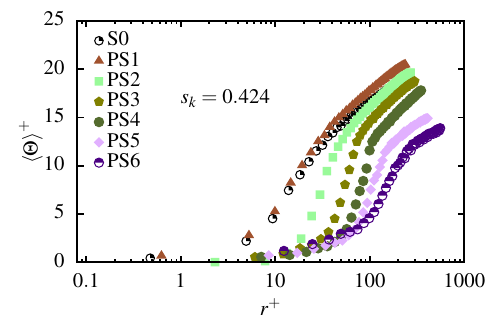}}

\subfloat[]{
\includegraphics[width=0.95\linewidth]{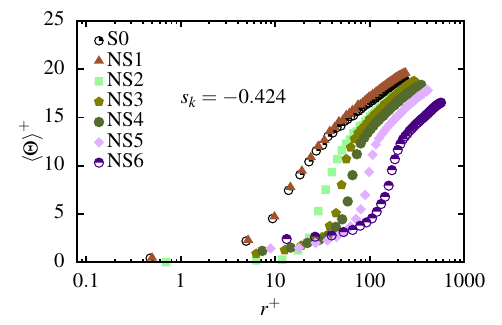}}
\caption{Mean temperature profile for (a) positively skewed and (b) negatively skewed surfaces in the inner wall units.}
\label{fig: Mean temperature profile}
\end{center}
\end{figure}

In Figs.~\ref{fig: Mean temperature profile}(a) and \ref{fig: Mean temperature profile}(b) mean temperature profiles, $\Theta^+$, are shown for all surfaces with $s_k=0.424$ and $s_k=-0.424$, respectively. Similar to $\left< u_x^+\right>$, an increasing downward shift is observed for $\Theta^+$ with increasing wall roughness height in the region $r^+ \geq 20$, while a slight upward shift is observed for $r^+ \leq 15$, signifying increased heat transfer close to the surface. For small roughness ($R_a/D=0.001$, $k_s^+=1$), the $\Theta^+$ profiles coincide with the ones for S0 irrespective of the skewness, as expected. Furthermore, the downward shift in the $\left< u_x^+\right>$ profile is consistently larger than that of the $\Theta^+$ profile at the same $k_s^+$ value, consistently with previous DNS studies \cite{peeters2019turbulent,macdonald2019roughness,kuwata2021direct,forooghi2018systematic,garg2024heat}. The direct comparison of $\Theta^+$ profiles in Figs.~\ref{fig: Mean temperature profile}(a) and \ref{fig: Mean temperature profile}(b) for comparable $R_a/D$ value shows that peak-dominated surfaces ($s_k=0.424$) have larger impact on heat transfer enhancement than the valley-dominated ones ($s_k=-0.424$). In other terms, surfaces with the same value of $R_a/D$ but different $s_k$ lead to slightly different predictions of $k_s^+$, and hence, larger roughness height and peak-dominated surfaces result in higher turbulence levels and eventually higher heat transfer rates. 

\begin{figure}[!ht]
\begin{center}
\subfloat[]{
\includegraphics[width=0.95\linewidth]{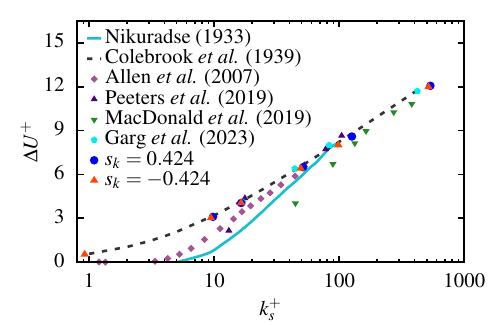}}

\subfloat[]{
\includegraphics[width=0.95\linewidth]{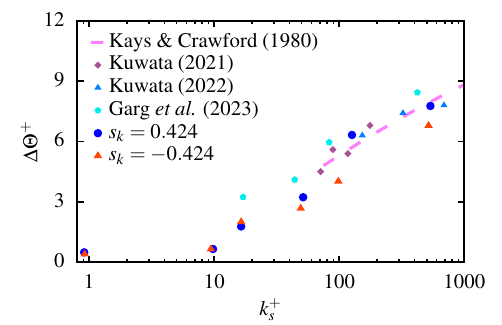}}
\caption{(a) Roughness function, $\Delta U^+$ and (b) temperature difference, $\Delta \Theta^+$, as a function of equivalent sand-grain roughness Reynolds number, $k_s^+$, for positively skewed (blue circles) and negatively skewed (red triangles) surfaces.}
\label{fig: Roughness function}
\end{center}
\end{figure}

The magnitudes of the downward shift can be measured by using the roughness function ($\Delta U^+$) and temperature difference function ($\Delta \Theta^+$). 
Figure \ref{fig: Roughness function}(a) shows $\Delta U^+$ and Fig. \ref{fig: Roughness function}(b) shows $\Delta \Theta^+$ as a function of $k_s^+$, indicating the changes in the mean profiles due to wall roughness. The results for $s_k=0.424$ are shown in blue circles and the ones with $s_k=-0.424$, in red triangles. The chosen scaling ensures that the roughness function collapses with Colebrook and Nikuradse's sand-grain data in the fully rough regime, as detailed by Garg \textit{et al.}~\cite{garg2023large,garg2024large}. It is important to note that $k_s$ must be dynamically determined for each specific rough surface and does not represent a simple geometric length scale of the roughness. Consequently, in this study, the value of $k_s^+$ is not known in advance and has been computed by extracting $\Delta U^+$ from Fig. \ref{fig: Mean flow profile}(a) for positively skewed surfaces and Fig. \ref{fig: Mean flow profile}(b) for negatively skewed surfaces and then using a Colebrook-type roughness function, $\Delta U^+ = 2.44\log{\left(1+0.26k_s^+\right)}$~\cite{colebrook1939correspondence}. 

Figure~\ref{fig: Roughness function}(a) illustrates the progressive increase of $\Delta U^+$ with $k_s^+$, eventually approaching the asymptotic limit for the fully rough regime. The fully rough regime for the present rough surface is observed in the range of $80 < k_s^+ < 537$, which is somewhat comparable to sand-grain roughness with $k_s^+ \simeq 70$ \cite{nikuradse1950laws}. The literature reveals similar roughness functions obtained for sand-grain roughness at approximately the same $Re$ number, even though the roughness heights significantly differ \cite{busse2017reynolds,thakkar2017surface,barros2019characteristics,busse2020influence,garg2023large,garg2024heat}. Interestingly, for $k_s^+ = 1$, peak-dominated (PS1) and valley-dominated (NS1) surfaces result in the same $\Delta U^+$ values, as seen from $\left<u^+\right>$ profiles. With increasing $k_s^+$ values, we found that surfaces with $s_k > 0$ result in a larger value of $\Delta U^+$ compared to the one with $s_k <0$ even if the surfaces share the same $R_a/D$ values. This suggests that the roughness height alone is inadequate for scaling the momentum and heat transfer deficit resulting from surface roughness. 

Similarly, the temperature difference function ($\Delta \Theta^+$) is computed and shown in Fig.~\ref{fig: Roughness function}(b) for $s_k > 0$ (blue circles) and $s_k < 0$ (red triangles). Globally, $\Delta \Theta^+$ initially increases with $k_s^+$, but the rate of increase diminishes as $k_s^+$ becomes larger. This behavior aligns reasonably well with Kays and Crawford's correlation~\cite{kays1980convective} (dashed pink line): $\Delta \Theta^+ = \frac{Pr_t}{\kappa} \log(k_s^+)-3.48\frac{Pr_t}{\kappa} -1.25{k_s^+}^{0.22}Pr^{0.44}+\beta(Pr),$
where $\kappa=0.4$ is the von Kármán constant and $\beta(Pr)=5.6$ represent the log-law intercept. Additionally, we compare the results of $\Delta \Theta^+$ for different rough surfaces, including three-dimensional irregular roughness~\cite{kuwata2021direct,kuwata2022reynolds} with positive and negative $s_k$ values and three-dimensional AM roughness~\cite{garg2024heat} with positive $s_k$ values. In the limit of small $k_s^+ \leq 10$, the value of $\Theta^+ \approx 0$ indicates negligible impact of roughness height and $s_k$ on heat transfer enhancement. 
Interestingly, for $k_s^+ > 10$, we observe an increasing trend of $\Delta \Theta^+$ against $k_s^+$, which is found to be consistent with the reference data~\cite{kuwata2022reynolds,kuwata2021direct,garg2024heat}, despite variations in roughness type, $R_a$, and $s_k$ values from previous studies. This suggests that topological parameters, such as roughness type (AM or artificial), $s_k$, and $ES$, have little impact on the $\Delta \Theta^+$ trend against $k_s^+$, especially in the fully rough regime. However, differences in the absolute value of $\Theta^+$ are noticeable for AM roughness of Garg \textit{et al.}~\cite{garg2024heat}, potentially due to variations in $Re_b$ and $s_k$ values used in their simulations. 
Given that $\Delta U^+$ and $\Delta \Theta^+$ reflect enhancements of the momentum and heat transfer, respectively, the observation of $\Delta U^+ > \Delta \Theta^+$ indicates that the wall roughness increases the momentum transfer more than the heat transfer.

\subsection{Effective turbulent Prandtl number}
\begin{figure}[!ht]
\begin{center}
\subfloat[]{
\includegraphics[width=0.95\linewidth]{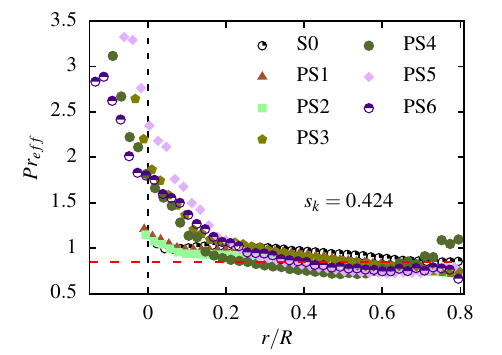}}

\subfloat[]{
\includegraphics[width=0.95\linewidth]{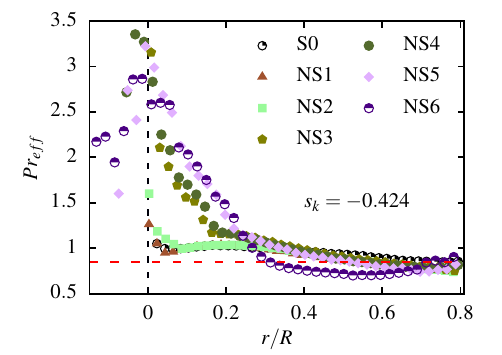}}

\caption{Effective Prandtl number plotted in outer units for (a) positively skewed surfaces and (b) negatively skewed surfaces. Results for the benchmark case, i.e., smooth surface, are also included for reference. The dashed red line shows $Pr_t=0.85$ fixed in the simulations.}
\label{fig: effective Prandtl}
\end{center}
\end{figure}

\begin{figure*}[!htb]
\begin{center}
\includegraphics[width=1\linewidth]{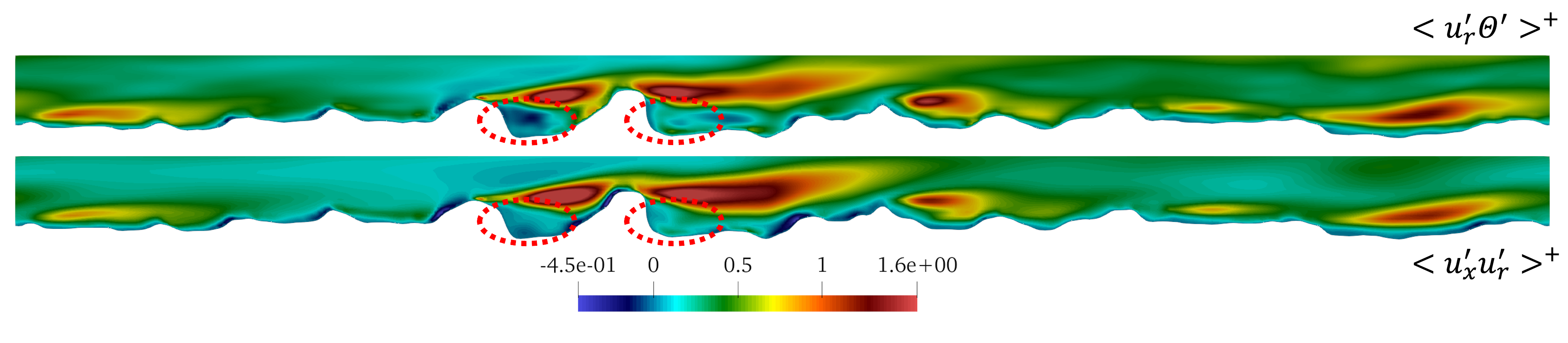}
\caption{Contour maps of wall-normal heat flux component, $\left< u'_r\Theta'\right>^+$, and Reynolds shear stress, $\left<u'_xu'_r \right>^+$, in a $x-y$ plane for $R_a/D=0.028$ and $s_k=0.424$. Red dashed lines indicate the region where the difference between $\left< u'_r\Theta'\right>^+$ and $\left<u'_xu'_r \right>^+$ is apparent.}
\label{fig: heat fluxes visualization}
\end{center}
\end{figure*}

According to Kader~\cite{kader1981temperature} the log-region of the mean temperature profile $\Theta$ in the absence of wall roughness can be described as
\begin{equation}
    \Theta^+(r^+)=\frac{Pr_t}{\kappa}\log(r^+)+\beta(Pr).
    \label{Eq: temperature profile}
\end{equation}
In this expression, $\beta(Pr) = \left( 3.85Pr^{1/3}-1.3\right)^2+\left(Pr_t/\kappa \right)\log \left(Pr\right)$. The literature typically suggests $Pr_t$ values in the range of $0.85$ to $1$. Spalart and Strelets~\cite{spalart2000mechanisms} proposed a method for determining an effective Prandtl number within a recirculation bubble as $Pr_{eff} = \nu_ {eff}/\alpha_{eff}$, where $\nu_{eff} = -\left<{u_i'u_j'}\right> \left<S_{ij}\right>$ and $\alpha_{eff} = -\left<{u_i'\Theta'}\right> \left< \partial{\Theta}/\partial x_i\right> / \left(\left< \partial {\Theta}/\partial x_i\right>\right)^2$. Far from the wall, $Pr_{eff} \approx Pr_t$. Profiles of $Pr_{eff}$ are displayed in Fig.~\ref{fig: effective Prandtl} for positively and negatively skewed surfaces. $Pr_{eff}$ shows similarity in cases such as S0, PS1, PS2, NS1, and NS2, where $k_s^+ < 10$. However, for larger $k_s^+$ values, $Pr_{eff}$ demonstrates a peak just below the plane $y/\delta=0$, which is equally pronounced for both positively and negatively skewed surfaces. This peak signifies a reduced effective thermal diffusivity, suggesting that regions $r/R < 0$ act as a thermal resistance. Moving away from the wall, $Pr_{eff}$ decreases with wall distance, reflecting the unmixedness of the scalar, as discussed in more detail in \cite{guezennec1990structure,abe2017relationship}. Additionally, far from the wall, simulations with roughness show a slightly smaller $Pr_{eff}$ compared to $Pr_t$ in smooth channel cases, especially for high $k_s^+$ with $s_k=-0.424$. These results indicate that roughness indeed influences $Pr_t$, with the influence of $s_k$ on these outcomes deemed insignificant for the surfaces examined in this study. Moreover, the reason for the rise in $Pr_{eff}$ within the rough wall can also be found in the contour maps of Reynolds shear stress, $\left< u'_xu_r'\right>^+$ and wall-normal component of heat flux, $\left< u'_r \Theta' \right>^+$ in Fig.~\ref{fig: heat fluxes visualization}. From Fig.~\ref{fig: heat fluxes visualization}, it is evident that  $\left< u'_r \Theta' \right>^+$ and $\left< u'_xu_r'\right>^+$  exhibit a similar trend but there is a noticeable difference just behind the roughness crests, where $\left< u'_r\Theta'\right>^+$ is somewhat smaller than $\left< u'_xu_r'\right>^+$. This is also reported by Kuwata~\cite{kuwata2021direct}, attributing it to recirculation zones behind the roughness crest, which had a detrimental effect on heat transfer. Moreover, the abrupt fluctuations in $Pr_{eff}$ in the vicinity of the bottom ($r/R < 0.01$) are associated with the effective eddy diffusivity $\nu_{eff}$. In this region, the local equilibrium state of turbulence breaks down; enhanced pressure diffusion acts as an energy source near the bottom of the rough surface, leading to an enhancement of the wall-normal Reynolds stress and the Reynolds shear stress~\cite{yuan2014roughness}.

\begin{figure*}[!htb]
\begin{center}
\subfloat[$R_a/D=0.001$]{
\includegraphics[width=0.45\linewidth]{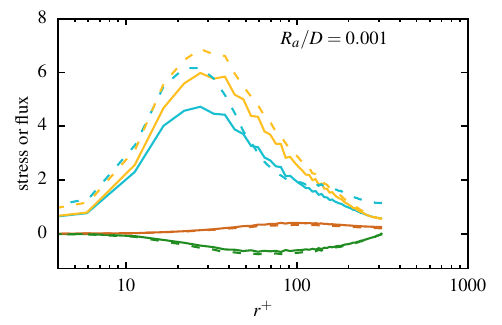}}
\subfloat[$R_a/D=0.006$]{
\includegraphics[width=0.45\linewidth]{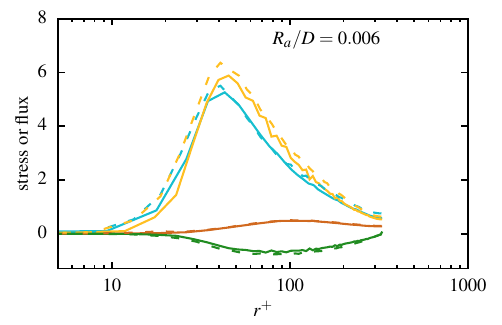}}

\subfloat[$R_a/D=0.012$]{
\includegraphics[width=0.45\linewidth]{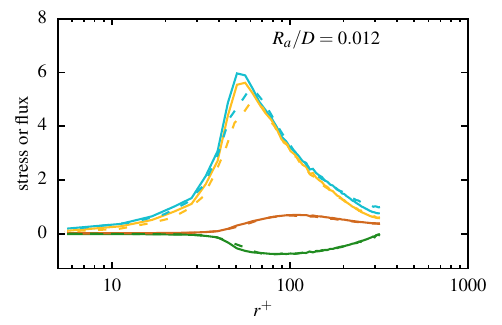}}
\subfloat[$R_a/D=0.015$]{
\includegraphics[width=0.45\linewidth]{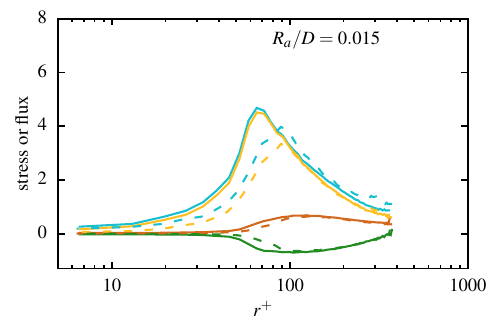}}

\subfloat[$R_a/D=0.020$]{
\includegraphics[width=0.45\linewidth]{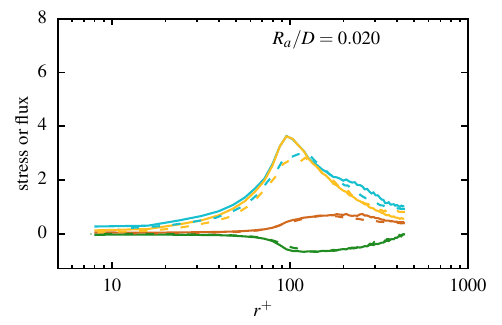}}
\subfloat[$R_a/D=0.028$]{
\includegraphics[width=0.45\linewidth]{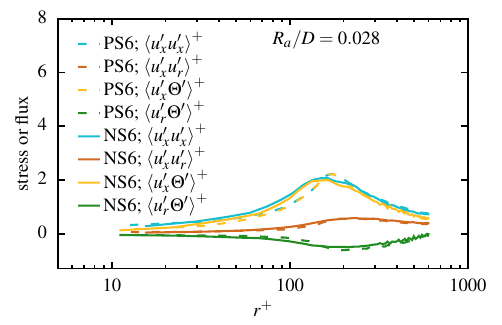}}
\caption{The wall-normal Reynolds shear stress, $\left<u'_x u'_r\right>$, streamwise velocity fluctuations, $\left<u'_xu'_x \right>$, normalized by $u_\tau^2$ and streamwise heat flux, $\left<u'_x \Theta' \right>$ and wall-normal heat flux, $\left<u'_r\Theta'\right>$, normalized by $u_\tau T_\tau$. 
Solid and dashed lines represent negatively skewed and positively skewed surface results, respectively.}  
\label{fig: Heat flux profiles}
\end{center}
\end{figure*}
\subsection{Impact of skewness and roughness height on Reynolds stresses and heat fluxes}
The temperature fluctuations, $\Theta'$, and velocity fluctuations, $u_x'$ and $u_r'$, give rise to significant mean Reynolds stresses and heat fluxes. Recent studies by Garg \textit{et al.}~\cite{garg2024heat} have shown the decreasing correlation between streamwise velocity fluctuations, $\left<u'_xu'_x\right>$ and streamwise component of heat flux, $\left<u'_x\Theta'\right>$, with increasing roughness height. However, the influence of roughness topologies, such as peak-dominated roughness with $s_k> 0$ and valley-dominated ones with $s_k < 0$, is still unclear. Thus, the effect of $s_k$ on the correlation between $\left<u'_x\Theta'\right>$ and $\left<u'_xu'_x\right>$ and between $\left<u'_r\Theta'\right>$ and $\left<u'_ru'_x\right>$ with increasing roughness height is investigated in this section. 

\begin{figure}[!htb]
\begin{center}

\includegraphics[width=0.95\linewidth]{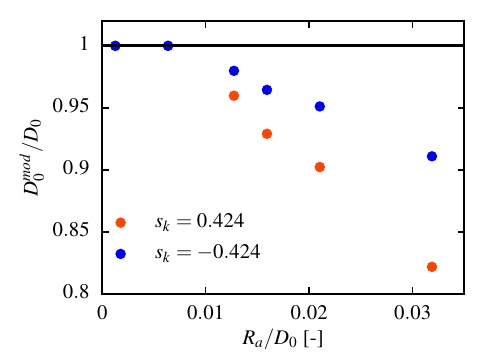}
\caption{Estimation of modified rough surface diameter due to negative velocities for positively and negatively skewed surfaces.}  
\label{fig: corrected diameter}
\end{center}
\end{figure}

In Fig.~\ref{fig: Heat flux profiles}, we compare $\left<u'_xu'_x\right>^+$ with $\left<u'_x\Theta'\right>^+$ and $\left<u'_xu'_r\right>^+$ with $\left<u'_r\Theta'\right>^+$, for the range of $R_a/D=0.001-0.028$, yielding a range of $k_s^+=1-536$. All solid and dashed lines represent negatively skewed and positively skewed surfaces, respectively. All cases show that irrespective of the roughness height and skewness values, the general trend between $\left<u'_r\Theta'\right>^+$ and $\left<u'_xu'_r\right>^+$ is practically identical, with a slight difference in the magnitude, $\left<u'_r\Theta'\right>^+$ being slightly smaller than $\left<u'_xu'_r\right>^+$. This is consistent with $Pr_{eff}$ being greater than one in the vicinity of the wall and the results shown in Fig.~\ref{fig: heat fluxes visualization}, indicating the recirculation zone behind the roughness wall acting as a thermal resistance. With an increase in roughness height, we observe that the peak value of these quantities shifts outwards after a critical value of $R_a/D > 0.006$, i.e., outside the hydraulically smooth regime. For larger $R_a/D$ values, the impact of $s_k$ starts to become visible for $R_a/D > 0.012$ as the peak location starts moving outward for the positively skewed surfaces compared to the negatively skewed ones, which means positively skewed surfaces generate larger friction, consistently with our earlier results. This outward shift can also be explained by calculating the modified diameter, $D^{mod}_0 = D_0-d$, where $D_0$ is the smooth pipe diameter. The estimation of normalized $D^{mod}_0$ is shown in Fig.~\ref{fig: corrected diameter}. Here we can see that the “effective” diameter is about 10\% larger for the negative skewness than the positive one (at $Ra/D_0 = 0.032$), having a profound influence on the overall flow. Effectively, the positive one squeezes the flow towards the center, which causes a larger outward shift of the peak for positively skewed surfaces compared to negative ones. A slight difference between $\left<u'_xu'_x\right>$ and $\left<u'_x\Theta'\right>$ is seen in the magnitude. The general trend is very similar for these two quantities as well. Interestingly, the difference between $\left<u'_x\Theta'\right>$ and $\left<u'_xu'_x\right>$ is a non-monotonic function of $R_a/D$. 

%
\subsection{Impact of skewness and roughness height on effective heat transfer}
\begin{figure*}[!htb]
\begin{center}
\subfloat[]{
\includegraphics[width=0.45\linewidth]{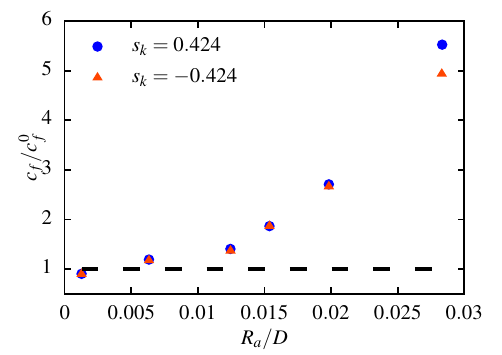}}
\subfloat[]{
\includegraphics[width=0.45\linewidth]{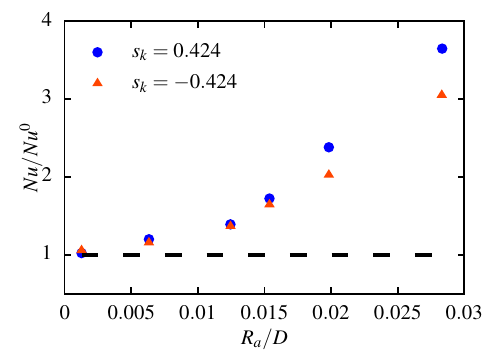}}

\subfloat[]{
\includegraphics[width=0.45\linewidth]{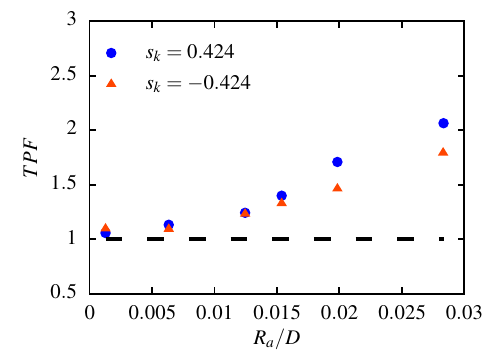}}
\caption{Estimation of (a) skin friction factor, $c_f$, (b) Nusselt number, $Nu$, and (c) thermal performance factor, TPF, as a function of $R_a/D$. All values are normalized by the benchmark case, i.e., the smooth pipe flow. The dashed black line represents the expected smooth pipe flow results.}
\label{fig: Global heat transfer}
\end{center}
\end{figure*}

Figure~\ref{fig: Global heat transfer}(a) presents the friction factor as a function of $R_a/D$ for various rough surfaces with $s_k=0.424$ and $s_k=-0.424$ values, normalized with respect to the numerical results obtained for smooth pipe flow using the Moody chart. The plots cover a $R_a/D$ number range of 0.001 to 0.028. Due to the lack of experimental validation for rough surfaces, we employed LES with the WALE model for smooth pipe flow at the same mesh resolution as rough pipes and matched Reynolds numbers. Previous studies by Garg \textit{et al.}~\cite{garg2024large,garg2024heat} utilized LES where the WALE model results for smooth pipe flow were used as a validation tool, showing excellent agreement with theoretical predictions from the Moody chart, with a deviation of less than 3\% (refer to Fig. 10(a) dashed black and solid green lines of Garg \textit{et al.}~\cite{garg2024heat}). This validation process instills confidence in the accuracy and robustness of our numerical approach for predicting the friction factor behavior in smooth pipe flow. By extension, it reinforces the reliability of our numerical results for turbulent flow over rough surfaces, given the utilization of the same numerical approach. 

In Fig.~\ref{fig: Global heat transfer}(a), the normalized $c_f/c^0_f$ values indicate a consistent upward trend with increasing roughness height ($R_a/D$ increases from 0.001 to 0.028), regardless of $s_k$ values. For PS1 and NS1 with $R_a/D = 0.001$ and $k_s^+ \approx 1 $, representing hydraulically smooth roughness, $c_f$ is negligibly impacted, leading to $c_f/c_f^0$ approaching 1. In the transition regime of PS2 to PS4 and NS2 to NS4, where $k_s^+$ ranges from 9 to 50, roughness elements disrupt the laminar flow, inducing additional drag and increasing resistance compared to smooth pipe flow, thus explaining the observed increasing behavior of $c_f/c_f^0$. The influence of $s_k$ remains insignificant up to $R_a/D=0.03$. However, as $R_a/D$ surpasses this threshold, rough surfaces PS5, PS6, NS5, and NS6 yield $k_s^+$ values between $100$ and $536$, indicating a fully rough regime where intensified turbulence significantly increases $c_f$. The impact of $s_k$ becomes pronounced, with positively skewed surfaces displaying higher $c_f$ than negatively skewed ones, indicating a greater imbalance between turbulence effects and roughness-induced drag in peak-dominated surfaces. Surprisingly, up to considerable $R_a/D$ values, the impact of $s_k$ on $c_f$ is negligible.

In Fig.~\ref{fig: Global heat transfer}(b), the normalized $Nu$ number is shown as a function of $R_a/D$ for various rough surfaces, normalized against the Nusselt number obtained for smooth pipe flow, $Nu^0$, using the Dittus-Boelter correlation $Nu^0=0.023Re_b^{0.8}Pr^{0.4}$. LES with the WALE model results for smooth pipe flow were utilized as a validation tool, exhibiting excellent agreement with theoretical predictions, with a deviation of less than 4\% (refer to Fig. 10(b) dashed black and solid green lines in Garg \textit{et al.}~\cite{garg2024heat}). Similar to the prediction of $c_f$, the observed behavior indicates potent heat transfer enhancement within the considered range of $R_a/D$ due to roughness-induced disturbances. In the hydraulically smooth regime, the impact of protrusions and valleys is negligible on the $Nu$ number, leading to $Nu/Nu^0$ approaching 1, regardless of $s_k$. As the transition roughness regime is entered, $Nu$ values begin to increase, resulting in potent heat transfer enhancement, with the impact of $s_k$ in this regime found to be negligible. However, in the fully rough regime, $Nu$ values exhibit a consistent upward trend, with the influence of $s_k$ becoming apparent. The results demonstrate that surfaces dominated by peaks ($s_k > 0$) result in higher heat transfer enhancement compared to those dominated by valleys ($s_k < 0$). 

To quantify the heat transfer enhancement effectiveness for the current roughness and assess the influence of surface topology, we calculated the thermal performance factor ($TPF$), representing the ratio of the relative change in heat transfer rate to the change in friction factor: $TPF = \left(Nu/Nu^0\right)/\left (c_f/c_f^0 \right ) ^{1/3}$, illustrated in Fig.~\ref{fig: Global heat transfer}(c). Three distinct regimes emerge. Firstly, the hydraulically smooth regime, where all rough surfaces perform equivalently to smooth ones. Secondly, the transitionally rough regime, where $TPF$ notably increases, remains between 1 and 1.5, signifying superior performance of all roughness configurations compared to smooth surfaces in transitional flow. The influence of $s_k$ becomes noticeable for $R_a/D = 0.015$. Lastly, in the fully rough regime, the impact of roughness height and skewness becomes markedly pronounced. Surfaces with positive $s_k$ values reveal a larger and more monotonic increase in $TPF$ compared to those with negative $s_k$ values. This highlights the critical importance of surface topology, where positively skewed surfaces outperform negatively skewed ones for similar roughness heights.

\subsection{Effect of surface anisotropy on the anisotropy of the Reynolds stress tensor}
The net anisotropy of the Reynolds stresses is commonly quantified using the second, $II_b$, and third, $III_b$, invariants of the normalized anisotropy tensor, $b_{ij}$, given by \cite{raupach1982averaging}
\begin{equation}
b_{ij} = \frac{\left< \overline{ u'_iu'_j}\right>}{\left< \overline{ u'_ku'_k}\right>} - \frac{1}{3} \delta_{ij}.
\label{anisotropy}
\end{equation}
The state of anisotropy can then be characterized with the two variables $\eta$ and $\xi$ defined as
\begin{equation}
\eta^2 = -\frac{1}{3}II_b
\end{equation}
and
\begin{equation}
\xi^3 = -\frac{1}{2}III_b.
\end{equation}

\begin{figure*}[!htb]
\begin{center}
\subfloat[S0]{
\includegraphics[width=0.372\linewidth]{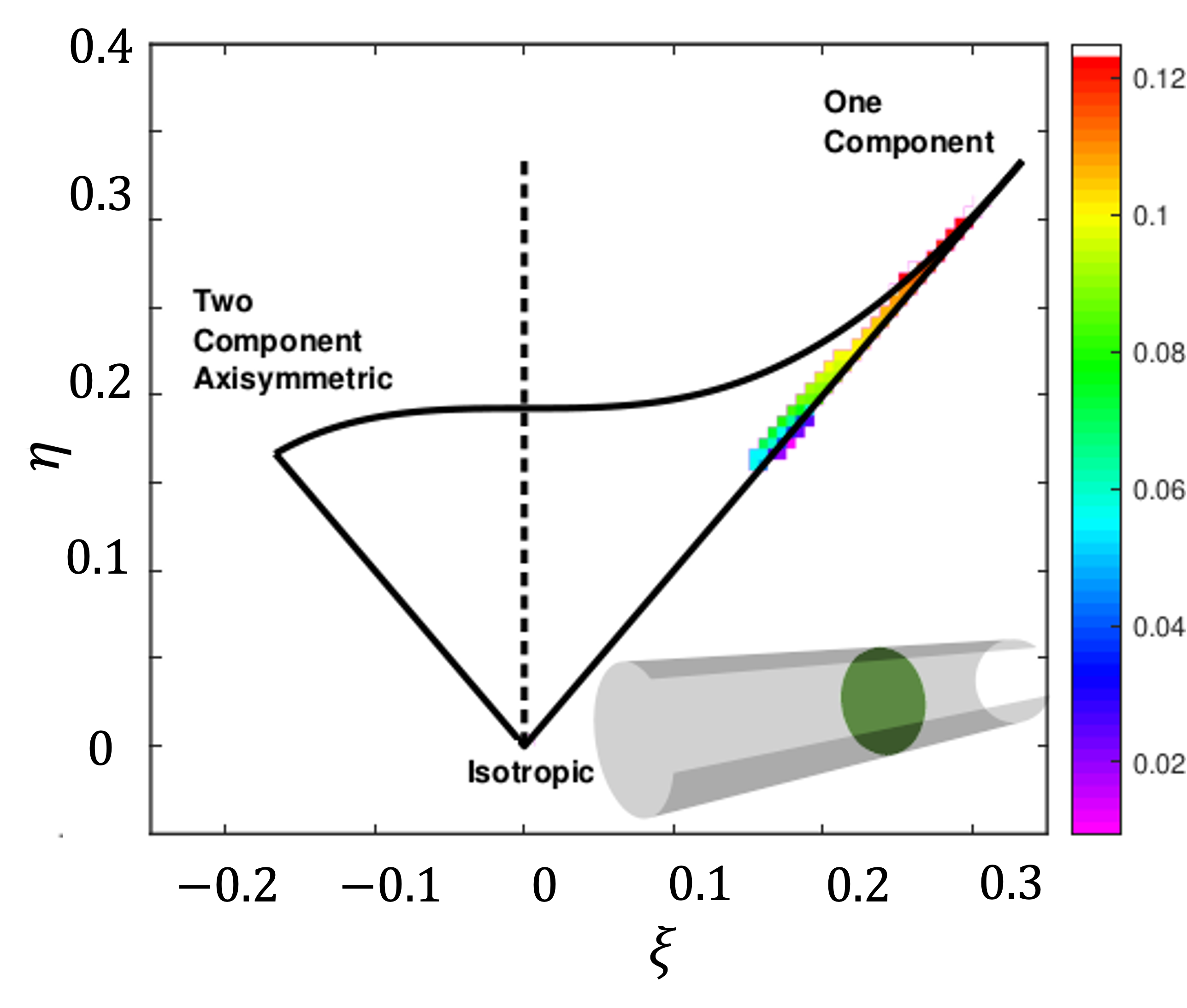}}
\subfloat[PS1]{
\includegraphics[width=0.35\linewidth]{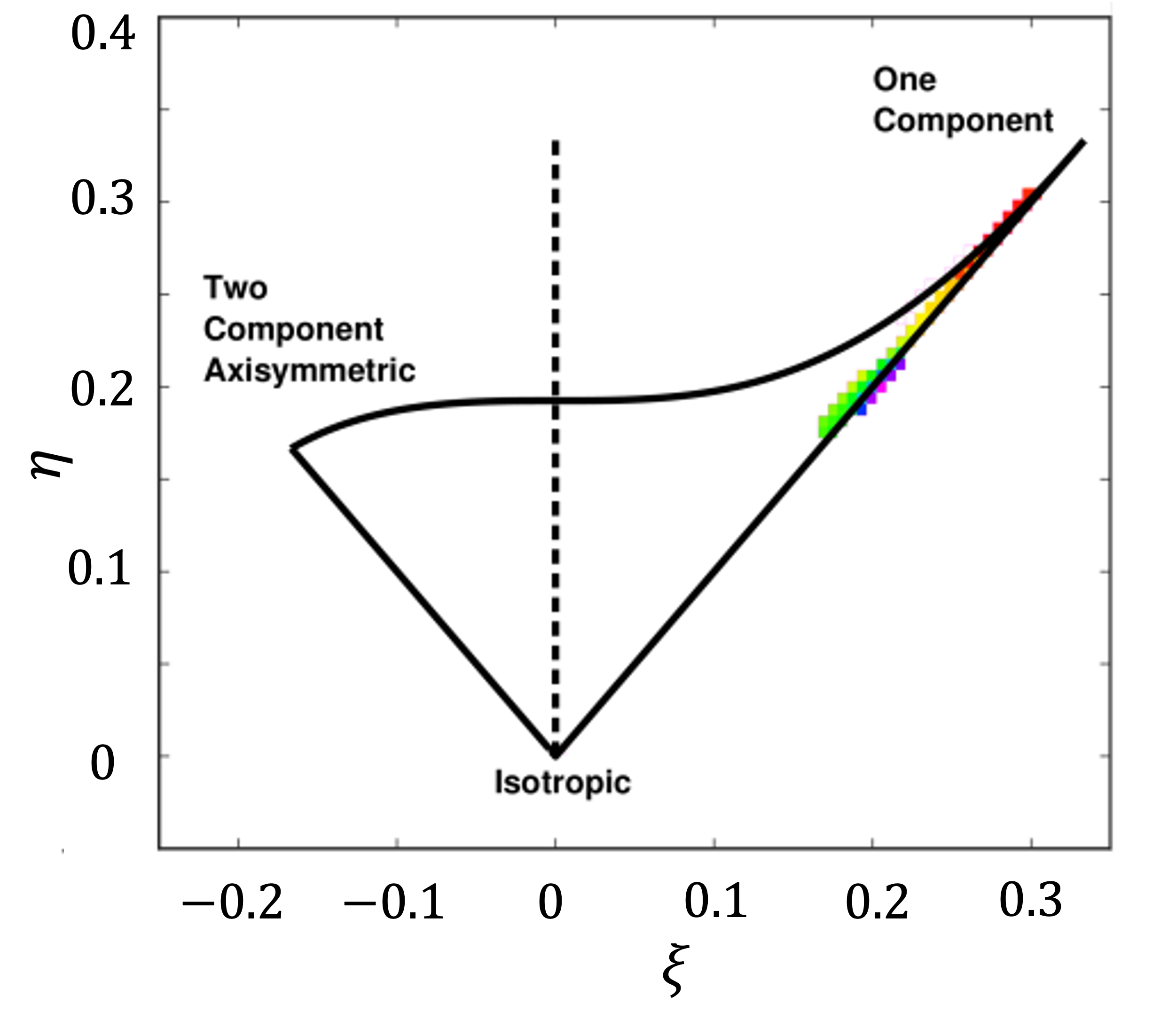}}

\subfloat[PS3]{
\includegraphics[width=0.35\linewidth]{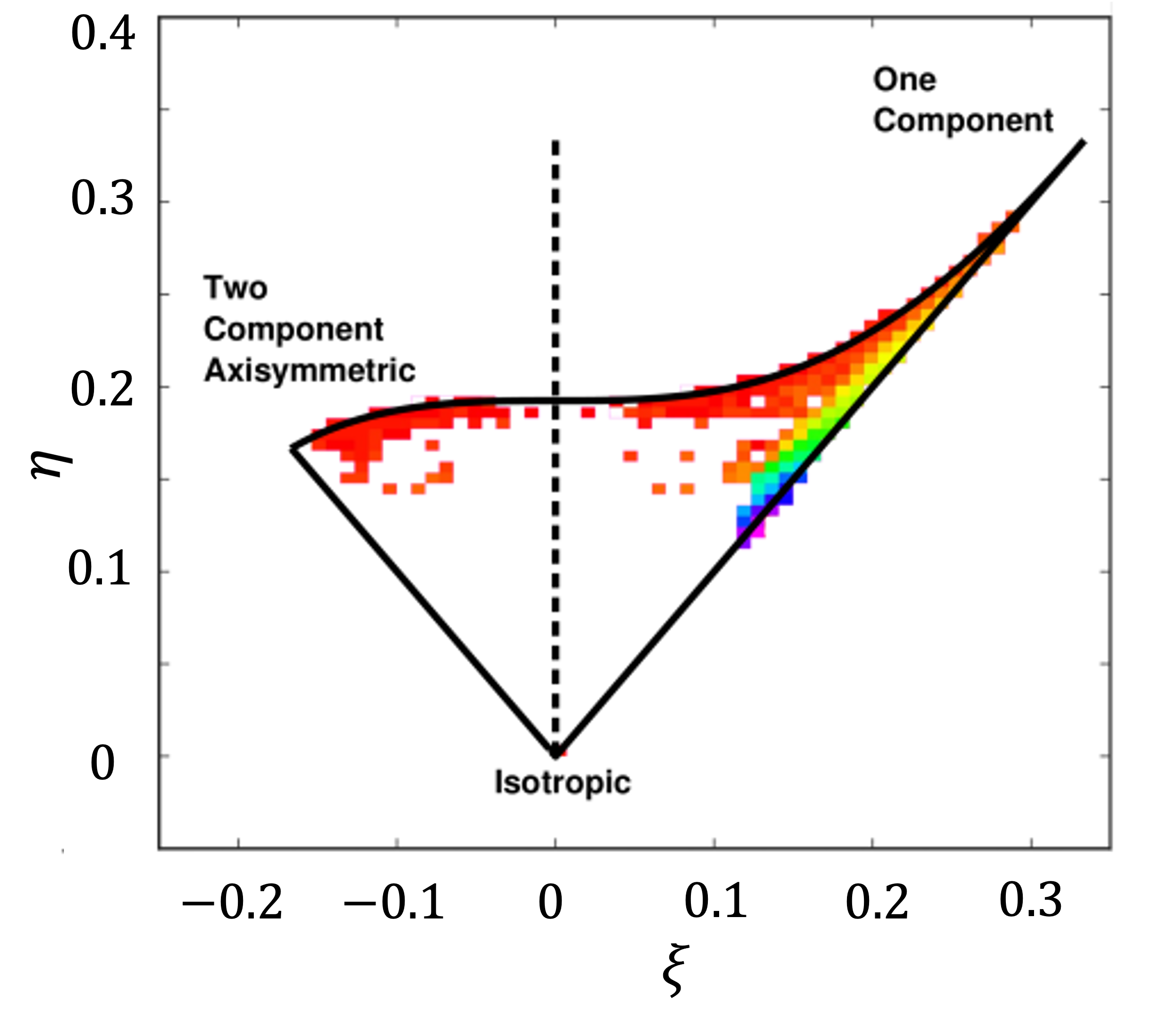}}
\subfloat[PS4]{
\includegraphics[width=0.35\linewidth]{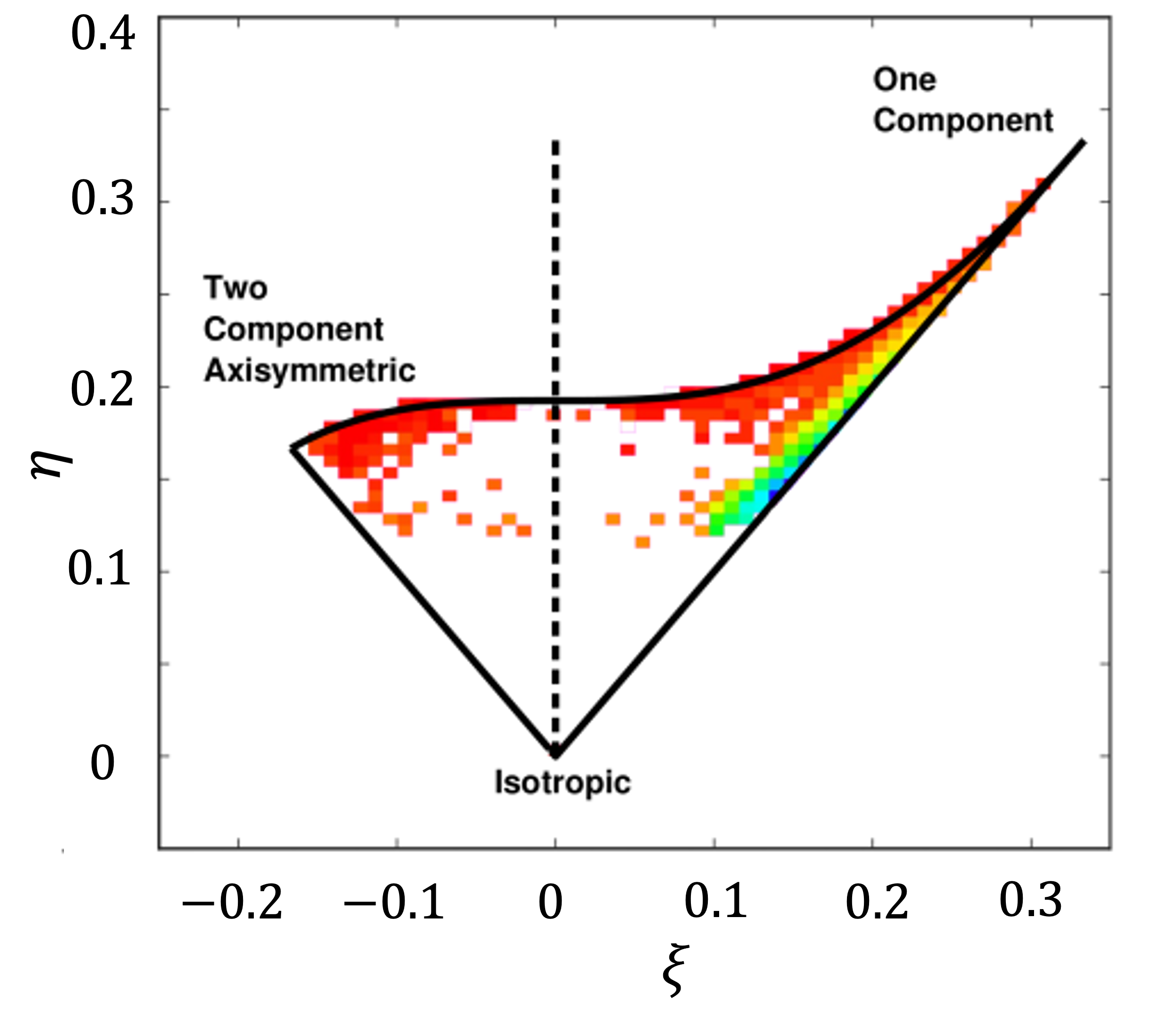}}

\subfloat[PS5]{
\includegraphics[width=0.35\linewidth]{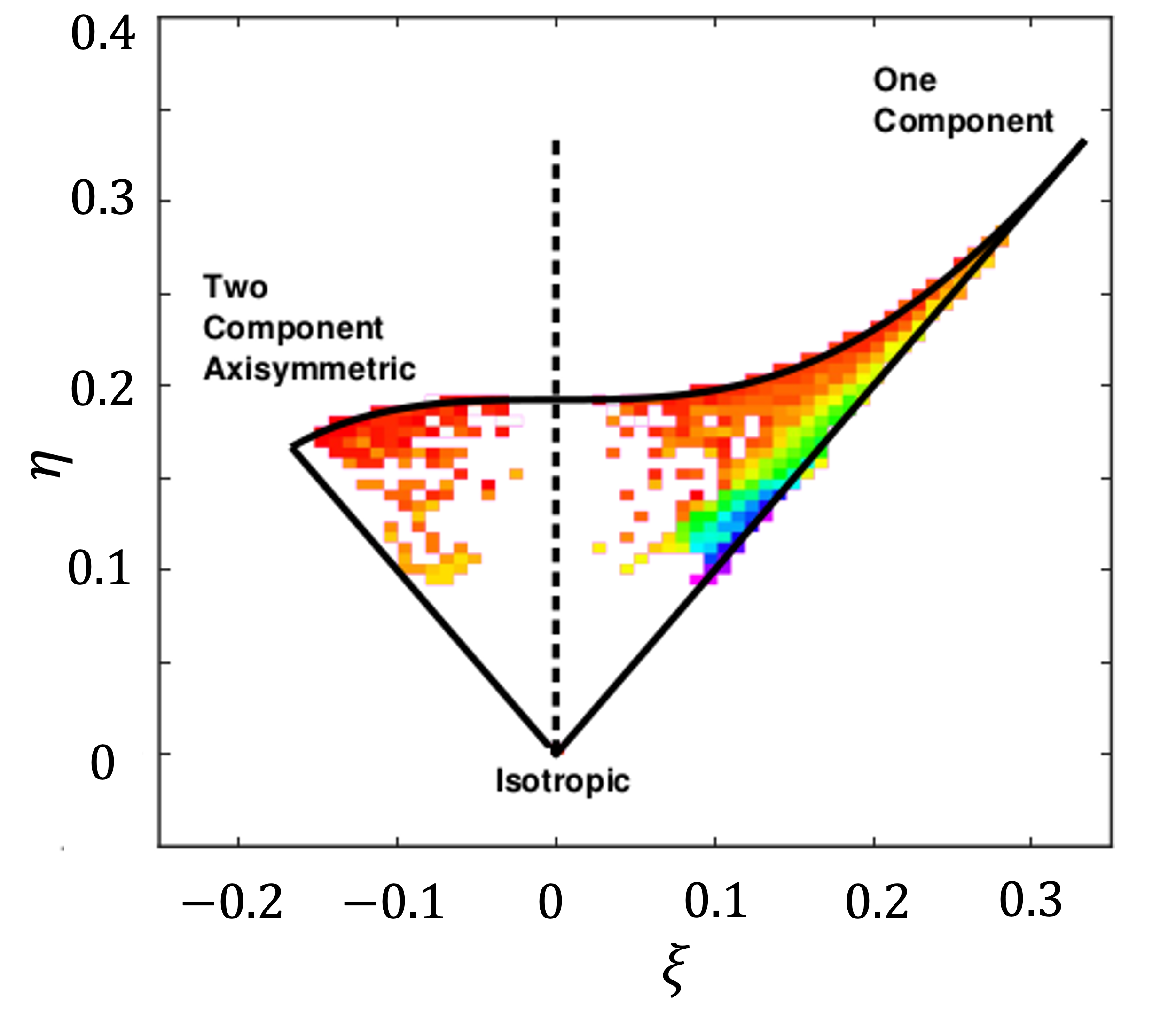}}
\subfloat[PS6]{
\includegraphics[width=0.35\linewidth]{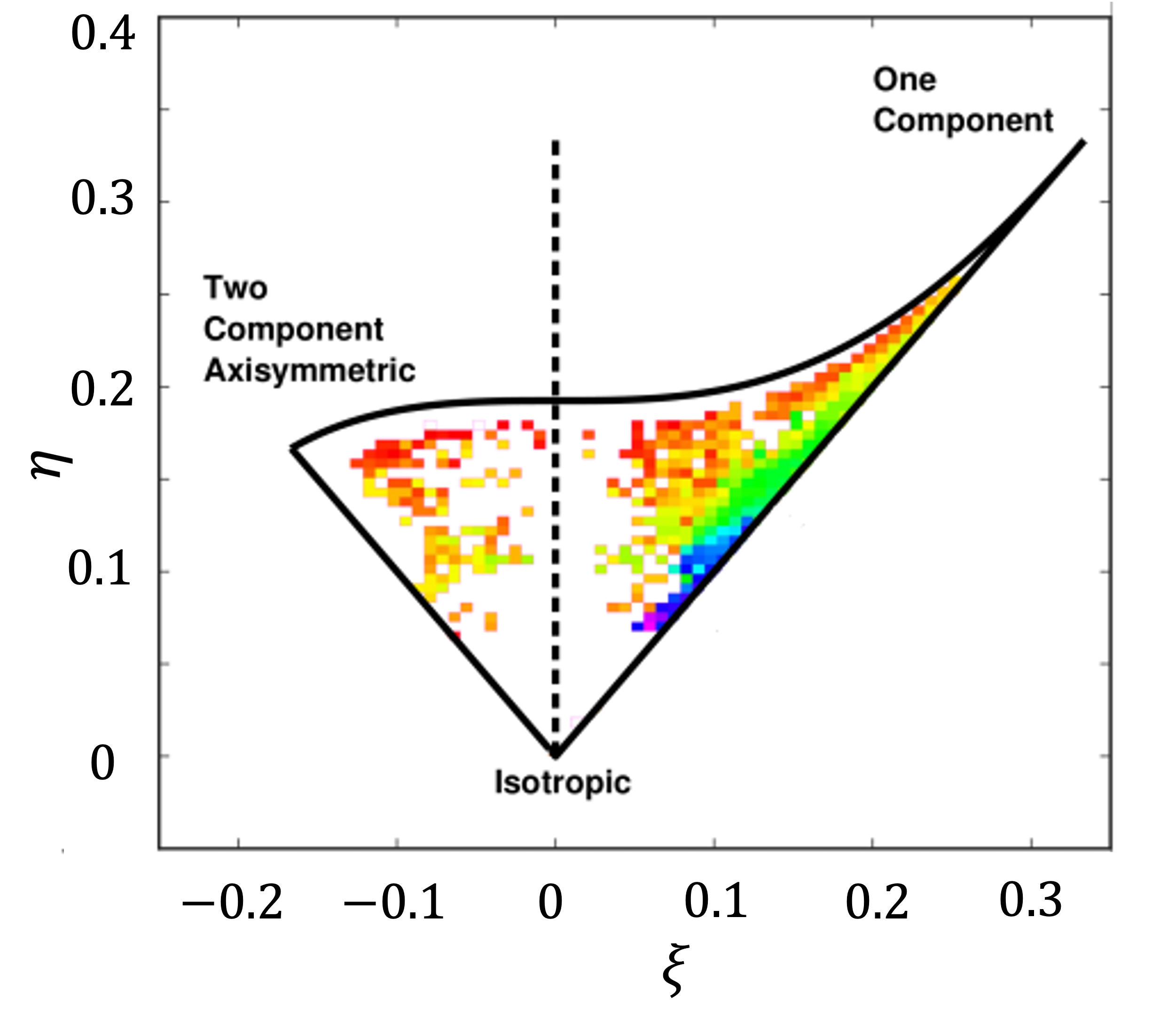}}
\caption{Anisotropy-invariant mapping of turbulence in (a) $R_a/D=0$ turbulent smooth pipe flow and rough pipe flow compiled from the present LES data at: (b) $R_a/D=0.001$, (c) $R_a/D=0.012$, (d) $R_a/D=0.015$, (e) $R_a/D=0.020$, (f) $R_a/D=0.028$, for positively skewed surfaces. The data points for each case are based on all cells in the domain at $x/D =4$ and colored with normalized wall distance values, $r^+$. Colormap varies from purple (minimum) to red (maximum).}
\label{fig: Lumley for PS}
\end{center}
\end{figure*}

\begin{figure*}[!htb]
\begin{center}
\subfloat[S0]{
\includegraphics[width=0.372\linewidth]{figures/lumleySliceS0-crop-crop.png}}
\subfloat[NS1]{
\includegraphics[width=0.35\linewidth]{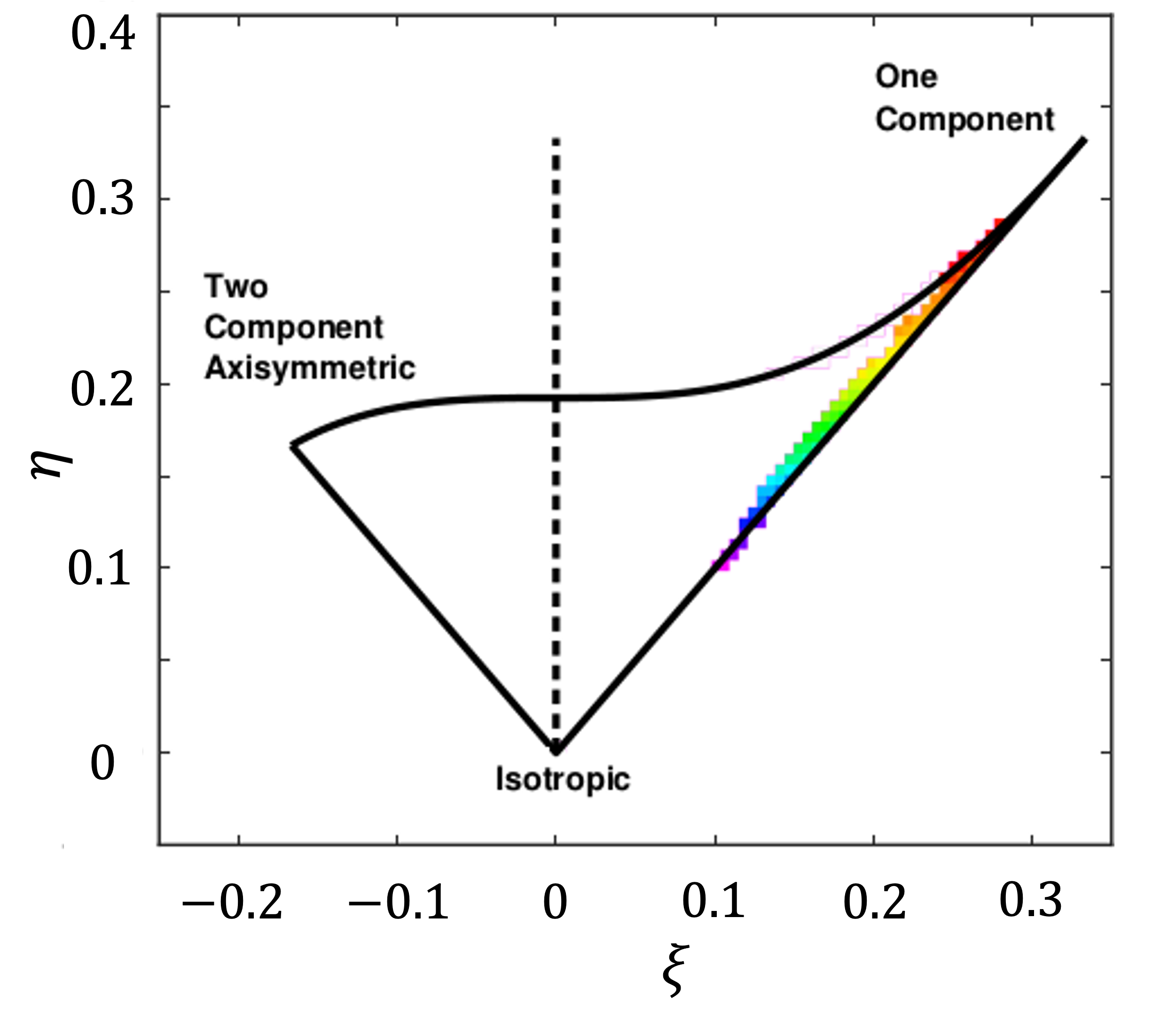}}

\subfloat[NS3]{
\includegraphics[width=0.35\linewidth]{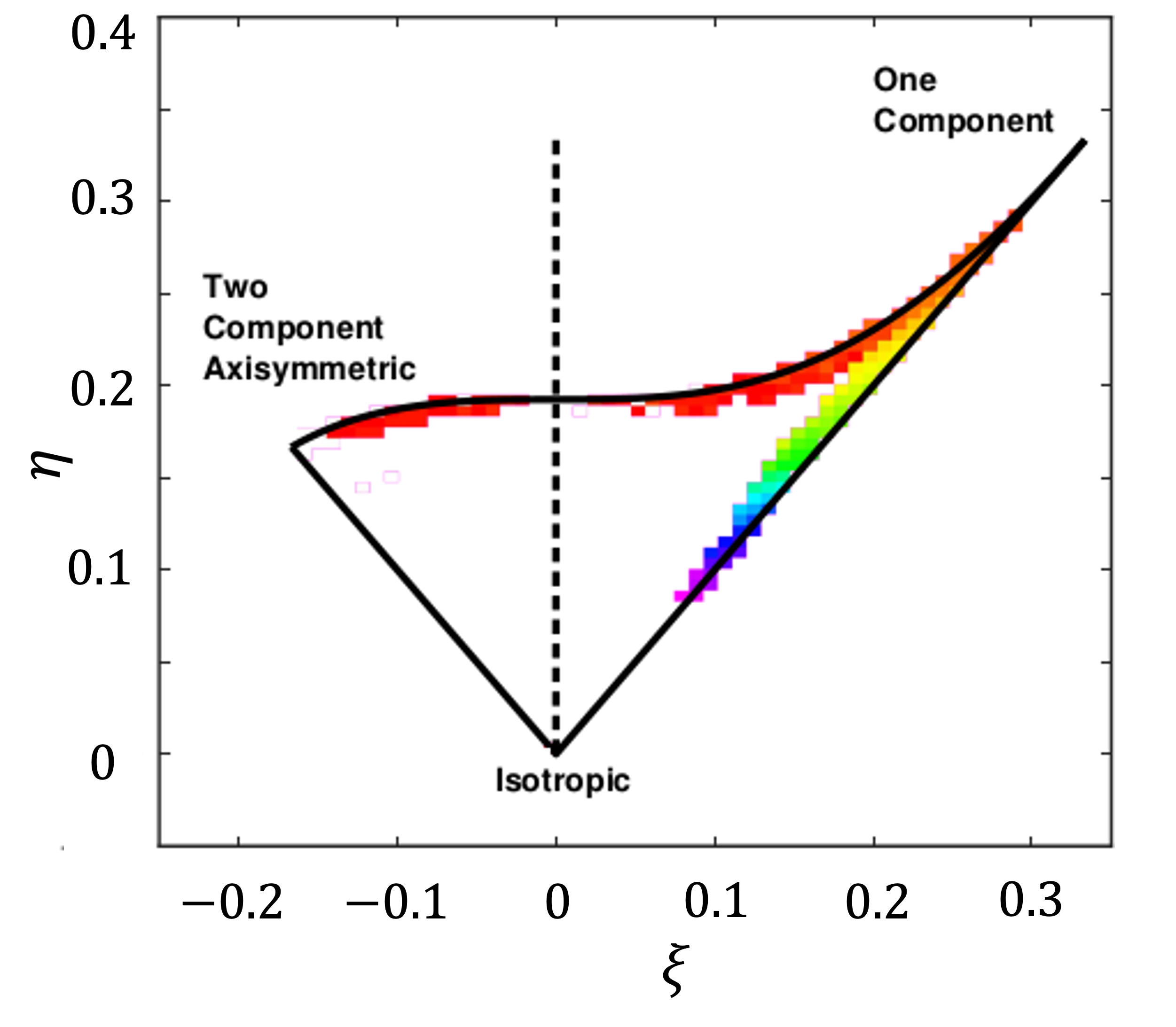}}
\subfloat[NS4]{
\includegraphics[width=0.35\linewidth]{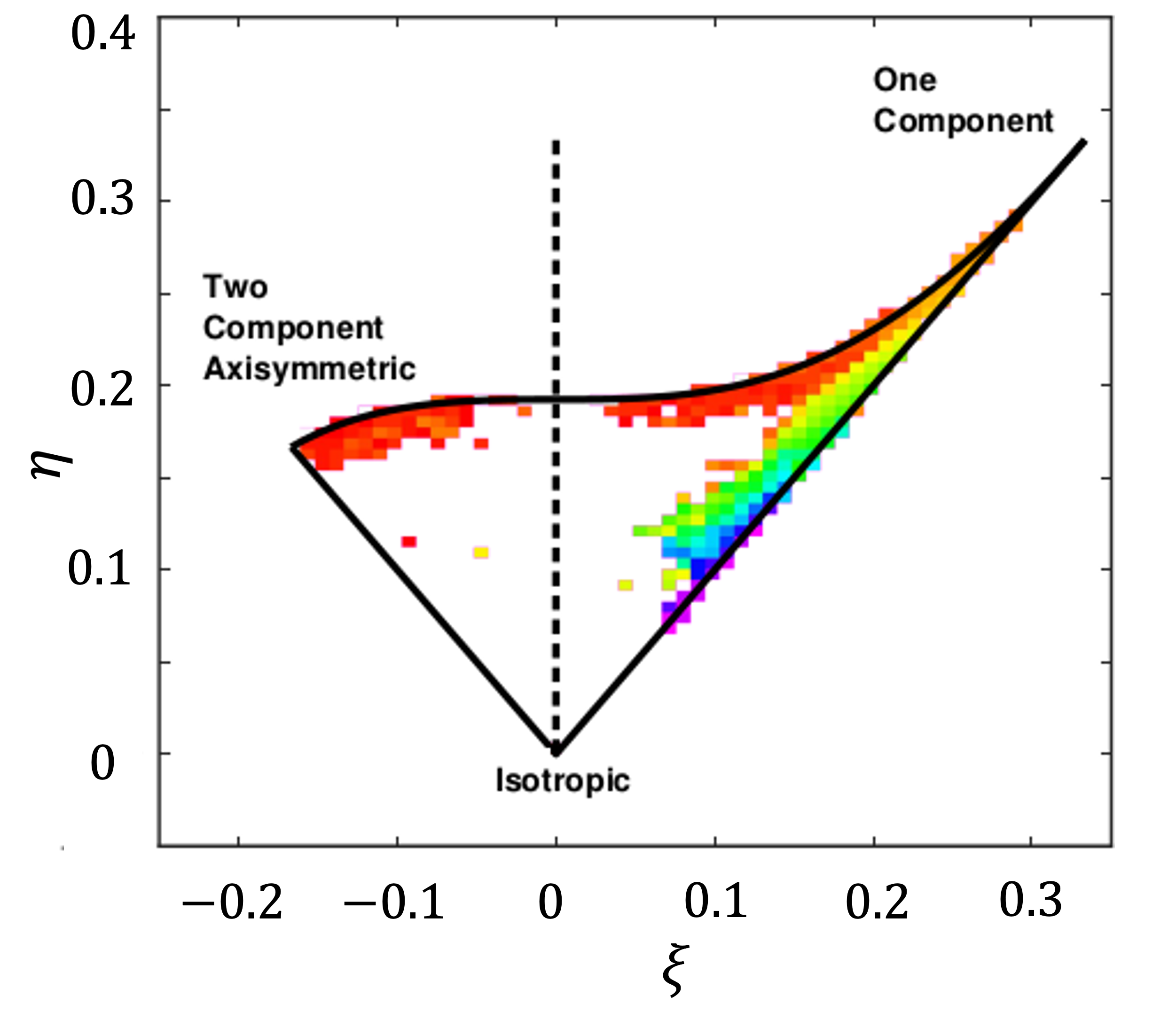}}

\subfloat[NS5]{
\includegraphics[width=0.35\linewidth]{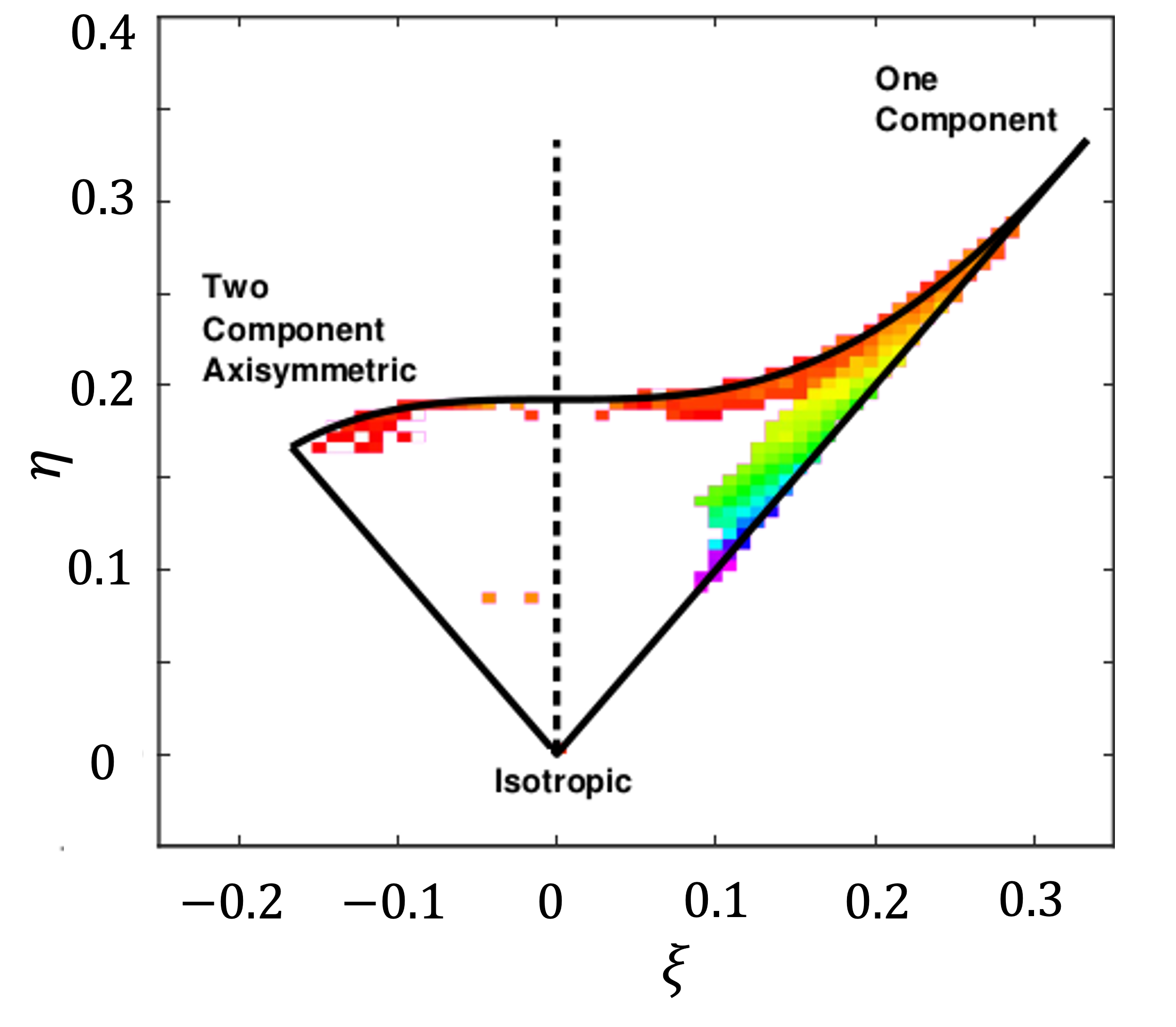}}
\subfloat[NS6]{
\includegraphics[width=0.35\linewidth]{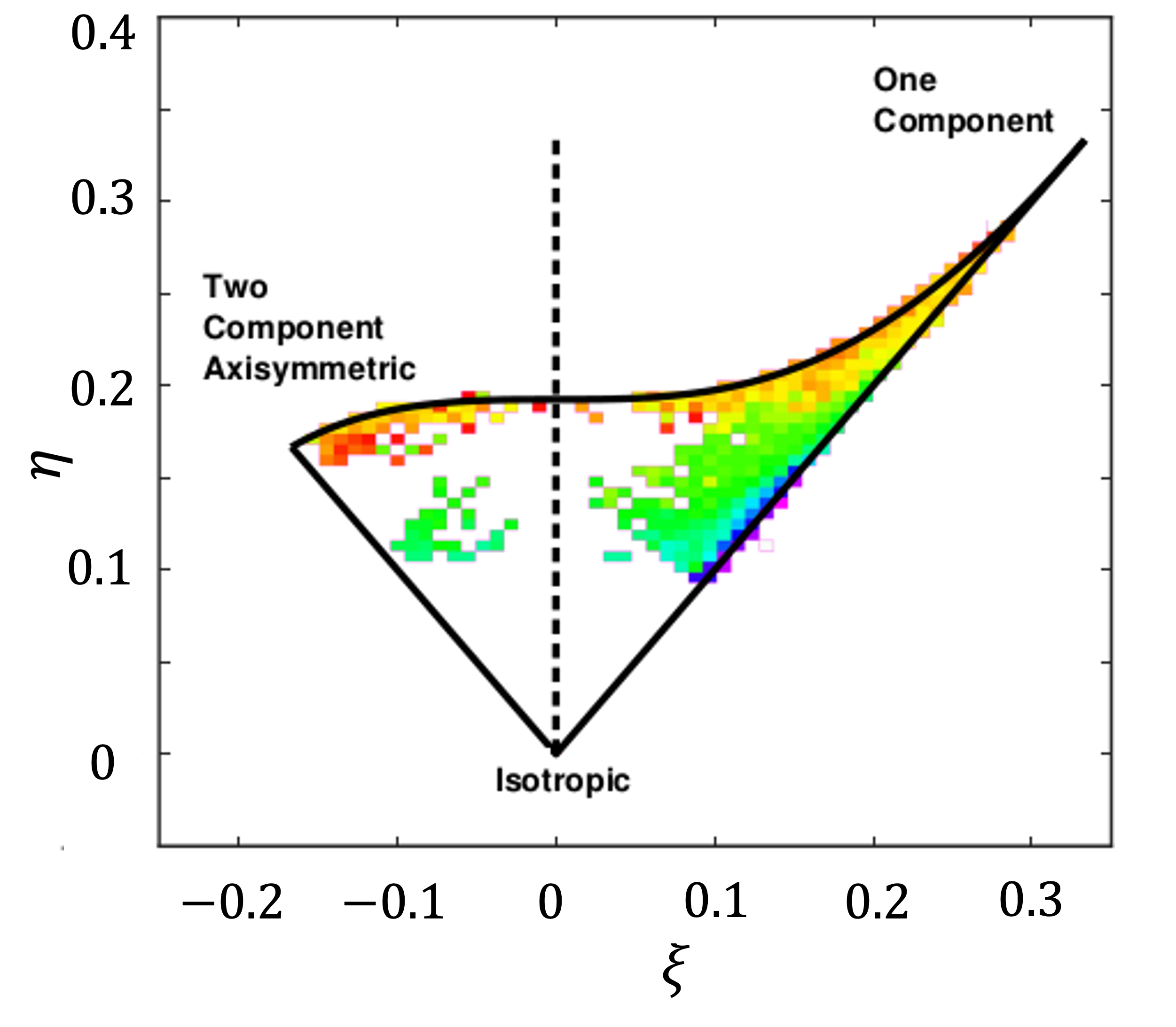}}
\caption{Anisotropy-invariant mapping of turbulence in (a) $R_a/D=0$ turbulent smooth pipe flow and rough pipe flow compiled from the present LES data at: (b) $R_a/D=0.001$, (c) $R_a/D=0.012$, (d) $R_a/D=0.015$, (e) $R_a/D=0.020$, (f) $R_a/D=0.028$, for negatively skewed surfaces. The data points for each case are based on all cells in the domain at $x/D =4$ and colored with normalized wall distance values, $r^+$. Colormap varies from purple (minimum) to red (maximum).}
\label{fig: Lumley for NS}
\end{center}
\end{figure*}

The Reynolds stress tensor's feasible states are confined within a triangular region in the ($\xi$, $\eta$) plane, known as the Lumley triangle. Distinct turbulence scenarios can be distinguished by examining the two invariants of $b_{ij}$ at the Lumley triangle's theoretical extremes. Garg \textit{et al.} \cite{garg2024large} provided a comprehensive analysis of these states. Figures~\ref{fig: Lumley for PS} and \ref{fig: Lumley for NS} depict Lumley triangle plots for rough surfaces with specific values of $s_k$ (0.424 and -0.424, respectively), along with a reference plot for smooth pipe flow (Figs.~\ref{fig: Lumley for PS}(a) and \ref{fig: Lumley for NS}(a)). Data are extracted from a $y-z$ plane slice at a consistent location for all surfaces, as illustrated in the inset of Fig.~\ref{fig: Lumley for PS}(a). Utilizing PDF post-processing, cell-centered values of $\xi$ and $\eta$ are visualized with a colormap to correspond with the wall-normal distance. In the smooth-wall turbulent pipe flow reference case, near-wall flow closely resembles a two-component state along the upper boundary of the Lumley triangle, progressing towards the one-component state at the triangle's upper-right apex. Maximum anisotropy occurs at approximately $r^+ \approx 8$, beyond which anisotropy diminishes. For $r^+>8$, the ($\xi$, $\eta$) curve aligns closely with the right boundary of the Lumley triangle, indicating proximity to an axisymmetric, rod-like state, but not fully reaching the maximum isotropy state at the triangle's base summit as $r^+$ increases. For smaller $R_a/D$ values, the behavior of surfaces PS1 and NS1 closely resembles that of smooth surface flows, irrespective of $s_k$ values.

For rough surfaces with $R_a/D > 0.01$, regardless of $s_k$ values, turbulent states occupy various positions within the Lumley triangle, except for the plain strain condition marked by the dashed line (see Figs. \ref{fig: Lumley for PS}(b-f) and \ref{fig: Lumley for NS}(b-f)). In the deepest valleys of the surfaces, the Reynolds stress anisotropy tensor tends toward a strongly anisotropic, one-component state. As $r^+$ increases, the results diverge from the trajectory observed for smooth-wall pipe flow on the $(\xi, \eta)$-map. Across all rough surfaces (PS1 to PS6 and NS1 to NS6), the flow converges towards the left side of the Lumley triangle, attaining an axisymmetric disk-like state at the roughness mean plane. Here, the streamwise and azimuthal Reynolds stresses exhibit comparable magnitudes. Anisotropy tends to center around the axisymmetric expansion and the two-component limit, indicating that one stress component predominates or that two components are similar in magnitude. Such axisymmetric, disk-like states of the Reynolds stress anisotropy tensor are characteristic of mixing layers. Similar behavior has been noted in turbulent flows over transverse bar roughness \cite{ashrafian2006structure}, k-type roughness \cite{smalley2002reynolds}, irregular roughness \cite{busse2020influence}, and recently in AM roughness~\citep{garg2023large}. Beyond the roughness mean plane, the trajectory shifts back towards the right side of the triangle, resembling an axisymmetric, rod-like state, mirroring the behavior of the smooth-wall case once the wall-normal coordinate surpasses the maximum roughness height. The most prevalent anisotropic states include axisymmetric expansion, one-component, two-component, and two-component axisymmetric states, with the likelihood of a specific turbulent state increasing with higher $R_a/D$.

The influence of $s_k$ values on Reynolds stress anisotropy is evident. For instance, comparing the PS4 surface in Fig. \ref{fig: Lumley for PS}(d) with the NS4 surface in Fig. \ref{fig: Lumley for NS}(d), both sharing the same $R_a/D$ but differing $s_k$ values, highlights this impact. The disparity in turbulent states between these surfaces stems from distinct flow interactions with surface irregularities. Surfaces characterized by positive and negative $s_k$ values represent two rough surface types: one dominated by peaks and the other by valleys, respectively. Peak-dominated surfaces exhibit protruding irregularities, promoting turbulence generation through flow acceleration, vorticity, and separation, resulting in a more intense and anisotropic turbulent flow. Conversely, valley-dominated surfaces have recessed features that dampen turbulence intensity by inducing flow deceleration and recirculation, resulting in a less intense and more isotropic turbulent flow. Thus, the physical mechanisms driving turbulence generation and anisotropy differ between peak-dominated and valley-dominated surfaces, explaining the observed differences in turbulent states.

\section{Conclusions}

We conducted a detailed examination of how wall roughness influences turbulent heat transfer in Additively Manufactured (AM) rough surfaces using roughness-resolved high-fidelity Large Eddy Simulations (LES) in OpenFOAM 7. We created six configurations of rough pipes from a single actual AM surface, maintaining fixed skewness and kurtosis while varying roughness height distributions. Additionally, we flipped these six surfaces to produce surfaces with fixed roughness height and kurtosis but with negative skewness. Precise spatial measurements at a constant bulk Reynolds number $Re_b=8000$ enabled us to numerically estimate the roughness function for all cases, which was then utilized to approximate the equivalent sand-grain roughness height, $k_s$. The temperature was treated as a passive scalar with a Prandtl number of 0.71, neglecting buoyancy effects. Consistent with our previous research~\cite{garg2023large,garg2024heat}, we found that wall roughness affects heat transfer and momentum differently. The temperature and momentum wall rough functions ($\Delta \Theta^+$, $\Delta U^+$) differed significantly, with the former being notably smaller than the latter. 
This discrepancy arises from high-temperature fluid from the bulk region penetrating the roughness layer, resulting in a larger wall-scaled mean temperature profile compared to the mean temperature profile, which is predominantly negative due to pressure effects within the roughness sublayer. Consequently, normalized temperature values in the bulk region are larger than the normalized mean velocity values. The discrepancy between $\Delta \Theta^+$ and $\Delta U^+$ directly challenges the validity of the Reynolds analogy under fully rough conditions, consistent with existing literature regardless of roughness nature, as the fact that $\Delta U^+ > \Delta \Theta^+$ can be attributed to the influence of pressure on the velocity field without a corresponding mechanism for the thermal field. This was further corroborated by the rapid increase in the effective Prandtl number ($Pr_{eff}$) within the rough wall, where the effective thermal diffusivity, due to combined turbulence and dispersion effects, is significantly smaller than the effective diffusivity within the rough wall. With increasing surface roughness height, $Pr_{eff}$ values increase near the roughness sublayer, indicating a reduction in the Reynolds analogy. The influence of skewness was found to be insignificant on the overall behavior. Furthermore, while the wall-normal Reynolds shear stress ($\left< u'_v u'_r\right>^+$) and heat flux ($\left<u'_r \Theta' \right>^+$) decreased with larger wall roughness height, their magnitudes remained similar for different values of surface roughness height ($R_a/D$). However, it is worth noting that the magnitude of $\left< u'_r \Theta'\right>^+$ was slightly smaller than $\left<u'_xu'_r\right>^+$, especially in the region just behind the roughness crest, highlighting the influence of recirculation bubbles on reducing heat transfer. The impact of skewness became more apparent for larger values of $R_a/D > 0.006$. For a fixed $R_a/D$ value, negatively skewed surfaces exhibited smaller $k_s^+$ values compared to positively skewed ones, indicating less turbulence in surfaces dominated by cavities and consequently less heat transfer. This was further quantified by visualizing turbulence states of $\left<u'_xu'_r\right>^+$, where peak-dominated surfaces demonstrated a higher probability of flow acceleration compared to valley-dominated ones. Evaluations of global features such as friction factor ($c_f/c_f^0$) and Nusselt number ($Nu/Nu^0$) demonstrated a clear dependence on surface roughness height and skewness. The influence of surface skewness was only evident on $c_f/c_f^0$ in the fully rough regime, while its effect on $Nu/Nu^0$ was noticeable in the transitional rough regime. Thermal performance evaluation further indicated that for efficient heat transfer enhancement, positively skewed surfaces outperformed negatively skewed ones in both transitional and fully rough flow regimes.

\section*{Acknowledgement}
The authors greatly appreciate the financial support provided by Vinnova under project number "2020-04529". Computer time was provided by the Swedish National Academic Infrastructure for Supercomputing (NAISS), partially funded by the Swedish Research Council through grant agreement no. 2018-05973. The authors also appreciate funding acquisition partners Setrab and Siemens Energy AB, and Sebastian Ritcher for conducting the surface roughness measurements. 
\section*{Author declarations}
\subsection*{Conflict of interest}
The authors have no conflicts of interest to disclose.
\subsection*{Data availability}
The data that support the findings of this study are available from the corresponding author upon reasonable request.
\newpage
\section*{References}

\providecommand{\noopsort}[1]{}\providecommand{\singleletter}[1]{#1}%
\begin{thebibliography}{64}%
\makeatletter
\providecommand \@ifxundefined [1]{%
 \@ifx{#1\undefined}
}%
\providecommand \@ifnum [1]{%
 \ifnum #1\expandafter \@firstoftwo
 \else \expandafter \@secondoftwo
 \fi
}%
\providecommand \@ifx [1]{%
 \ifx #1\expandafter \@firstoftwo
 \else \expandafter \@secondoftwo
 \fi
}%
\providecommand \natexlab [1]{#1}%
\providecommand \enquote  [1]{``#1''}%
\providecommand \bibnamefont  [1]{#1}%
\providecommand \bibfnamefont [1]{#1}%
\providecommand \citenamefont [1]{#1}%
\providecommand \href@noop [0]{\@secondoftwo}%
\providecommand \href [0]{\begingroup \@sanitize@url \@href}%
\providecommand \@href[1]{\@@startlink{#1}\@@href}%
\providecommand \@@href[1]{\endgroup#1\@@endlink}%
\providecommand \@sanitize@url [0]{\catcode `\\12\catcode `\$12\catcode
  `\&12\catcode `\#12\catcode `\^12\catcode `\_12\catcode `\%12\relax}%
\providecommand \@@startlink[1]{}%
\providecommand \@@endlink[0]{}%
\providecommand \url  [0]{\begingroup\@sanitize@url \@url }%
\providecommand \@url [1]{\endgroup\@href {#1}{\urlprefix }}%
\providecommand \urlprefix  [0]{URL }%
\providecommand \Eprint [0]{\href }%
\providecommand \doibase [0]{http://dx.doi.org/}%
\providecommand \selectlanguage [0]{\@gobble}%
\providecommand \bibinfo  [0]{\@secondoftwo}%
\providecommand \bibfield  [0]{\@secondoftwo}%
\providecommand \translation [1]{[#1]}%
\providecommand \BibitemOpen [0]{}%
\providecommand \bibitemStop [0]{}%
\providecommand \bibitemNoStop [0]{.\EOS\space}%
\providecommand \EOS [0]{\spacefactor3000\relax}%
\providecommand \BibitemShut  [1]{\csname bibitem#1\endcsname}%
\let\auto@bib@innerbib\@empty
\bibitem [{\citenamefont {Bunker}(2017)}]{bunker2017evolution}%
  \BibitemOpen
  \bibfield  {author} {\bibinfo {author} {\bibfnamefont {R.~S.}\ \bibnamefont
  {Bunker}},\ }\bibfield  {title} {\enquote {\bibinfo {title} {Evolution of
  turbine cooling},}\ }in\ \href@noop {} {\emph {\bibinfo {booktitle} {Turbo
  Expo: Power for Land, Sea, and Air}}},\ Vol.\ \bibinfo {volume} {50770}\
  (\bibinfo {organization} {American Society of Mechanical Engineers},\
  \bibinfo {year} {2017})\ p.\ \bibinfo {pages} {V001T51A001}\BibitemShut
  {NoStop}%
\bibitem [{\citenamefont {Adair}, \citenamefont {Kirka},\ and\ \citenamefont
  {Ryan}(2019)}]{adair2019additive}%
  \BibitemOpen
  \bibfield  {author} {\bibinfo {author} {\bibfnamefont {D.}~\bibnamefont
  {Adair}}, \bibinfo {author} {\bibfnamefont {M.}~\bibnamefont {Kirka}}, \ and\
  \bibinfo {author} {\bibfnamefont {D.}~\bibnamefont {Ryan}},\ }\bibfield
  {title} {\enquote {\bibinfo {title} {Additive manufacture of prototype
  turbine blades for hot-fired engine performance validation trials},}\ }in\
  \href@noop {} {\emph {\bibinfo {booktitle} {Turbo Expo: Power for Land, Sea,
  and Air}}},\ Vol.\ \bibinfo {volume} {58677}\ (\bibinfo {organization}
  {American Society of Mechanical Engineers},\ \bibinfo {year} {2019})\ p.\
  \bibinfo {pages} {V006T24A012}\BibitemShut {NoStop}%
\bibitem [{\citenamefont {Chowdhury}\ \emph {et~al.}(2022)\citenamefont
  {Chowdhury}, \citenamefont {Yadaiah}, \citenamefont {Prakash}, \citenamefont
  {Ramakrishna}, \citenamefont {Dixit}, \citenamefont {Gulta},\ and\
  \citenamefont {Buddhi}}]{Chowdhury2022LaserPB}%
  \BibitemOpen
  \bibfield  {author} {\bibinfo {author} {\bibfnamefont {S.}~\bibnamefont
  {Chowdhury}}, \bibinfo {author} {\bibfnamefont {N.}~\bibnamefont {Yadaiah}},
  \bibinfo {author} {\bibfnamefont {C.}~\bibnamefont {Prakash}}, \bibinfo
  {author} {\bibfnamefont {S.}~\bibnamefont {Ramakrishna}}, \bibinfo {author}
  {\bibfnamefont {S.}~\bibnamefont {Dixit}}, \bibinfo {author} {\bibfnamefont
  {L.~R.}\ \bibnamefont {Gulta}}, \ and\ \bibinfo {author} {\bibfnamefont
  {D.}~\bibnamefont {Buddhi}},\ }\bibfield  {title} {\enquote {\bibinfo {title}
  {Laser powder bed fusion: A state-of-the-art review of the technology,
  materials, properties \& defects, and numerical modelling},}\ }\href
  {https://api.semanticscholar.org/CorpusID:251530066} {\bibfield  {journal}
  {\bibinfo  {journal} {Journal of Materials Research and Technology}\ }
  (\bibinfo {year} {2022})}\BibitemShut {NoStop}%
\bibitem [{\citenamefont {Ngo}\ \emph {et~al.}(2018)\citenamefont {Ngo},
  \citenamefont {Kashani}, \citenamefont {Imbalzano}, \citenamefont {Nguyen},\
  and\ \citenamefont {Hui}}]{Ngo2018AdditiveM}%
  \BibitemOpen
  \bibfield  {author} {\bibinfo {author} {\bibfnamefont {T.~D.}\ \bibnamefont
  {Ngo}}, \bibinfo {author} {\bibfnamefont {A.~R.}\ \bibnamefont {Kashani}},
  \bibinfo {author} {\bibfnamefont {G.}~\bibnamefont {Imbalzano}}, \bibinfo
  {author} {\bibfnamefont {K.~T.}\ \bibnamefont {Nguyen}}, \ and\ \bibinfo
  {author} {\bibfnamefont {D.}~\bibnamefont {Hui}},\ }\bibfield  {title}
  {\enquote {\bibinfo {title} {Additive manufacturing (3d printing): A review
  of materials, methods, applications and challenges},}\ }\href
  {https://api.semanticscholar.org/CorpusID:139464688} {\bibfield  {journal}
  {\bibinfo  {journal} {Composites Part B: Engineering}\ } (\bibinfo {year}
  {2018})}\BibitemShut {NoStop}%
\bibitem [{\citenamefont {Colebrook}\ \emph {et~al.}(1939)\citenamefont
  {Colebrook}, \citenamefont {Blench}, \citenamefont {Chatley}, \citenamefont
  {Essex}, \citenamefont {Finniecome}, \citenamefont {Lacey}, \citenamefont
  {Williamson},\ and\ \citenamefont {Macdonald}}]{colebrook1939correspondence}%
  \BibitemOpen
  \bibfield  {author} {\bibinfo {author} {\bibfnamefont {C.~F.}\ \bibnamefont
  {Colebrook}}, \bibinfo {author} {\bibfnamefont {T.}~\bibnamefont {Blench}},
  \bibinfo {author} {\bibfnamefont {H.}~\bibnamefont {Chatley}}, \bibinfo
  {author} {\bibfnamefont {E.~H.}\ \bibnamefont {Essex}}, \bibinfo {author}
  {\bibfnamefont {J.~R.}\ \bibnamefont {Finniecome}}, \bibinfo {author}
  {\bibfnamefont {G.}~\bibnamefont {Lacey}}, \bibinfo {author} {\bibfnamefont
  {J.}~\bibnamefont {Williamson}}, \ and\ \bibinfo {author} {\bibfnamefont
  {G.~G.}\ \bibnamefont {Macdonald}},\ }\bibfield  {title} {\enquote {\bibinfo
  {title} {{Turbulent flow in pipes, with particular reference to the
  transition region between the smooth and rough pipe laws}},}\ }\href@noop {}
  {\bibfield  {journal} {\bibinfo  {journal} {J. Inst. Civ. Eng.}\ }\textbf
  {\bibinfo {volume} {12}},\ \bibinfo {pages} {393--422} (\bibinfo {year}
  {1939})}\BibitemShut {NoStop}%
\bibitem [{\citenamefont {Schlichting}(1937)}]{schlichting1937experimental}%
  \BibitemOpen
  \bibfield  {author} {\bibinfo {author} {\bibfnamefont {H.}~\bibnamefont
  {Schlichting}},\ }\href@noop {} {\emph {\bibinfo {title} {Experimental
  investigation of the problem of surface roughness}}},\ \bibinfo {number}
  {823}\ (\bibinfo  {publisher} {National Advisory Commitee for Aeronautics},\
  \bibinfo {year} {1937})\BibitemShut {NoStop}%
\bibitem [{\citenamefont {Nikuradse}(1933)}]{nikuradse1950laws}%
  \BibitemOpen
  \bibfield  {author} {\bibinfo {author} {\bibfnamefont {J.}~\bibnamefont
  {Nikuradse}},\ }\bibfield  {title} {\enquote {\bibinfo {title} {Laws of flow
  in rough pipes},}\ }\href@noop {} {\  (\bibinfo {year} {1933})}\BibitemShut
  {NoStop}%
\bibitem [{\citenamefont {Clauser}(1954)}]{clauser1954turbulent}%
  \BibitemOpen
  \bibfield  {author} {\bibinfo {author} {\bibfnamefont {F.~H.}\ \bibnamefont
  {Clauser}},\ }\bibfield  {title} {\enquote {\bibinfo {title} {Turbulent
  boundary layers in adverse pressure gradients},}\ }\href@noop {} {\bibfield
  {journal} {\bibinfo  {journal} {J. Aeronaut. Sci.}\ }\textbf {\bibinfo
  {volume} {21}},\ \bibinfo {pages} {91--108} (\bibinfo {year}
  {1954})}\BibitemShut {NoStop}%
\bibitem [{\citenamefont {Dittus}(1930)}]{dittus1930heat}%
  \BibitemOpen
  \bibfield  {author} {\bibinfo {author} {\bibfnamefont {F.~W.}\ \bibnamefont
  {Dittus}},\ }\bibfield  {title} {\enquote {\bibinfo {title} {Heat transfer in
  automobile radiators of the tubular type},}\ }\href@noop {} {\bibfield
  {journal} {\bibinfo  {journal} {Univ. of California Pub., Eng.}\ }\textbf
  {\bibinfo {volume} {2}},\ \bibinfo {pages} {443--461} (\bibinfo {year}
  {1930})}\BibitemShut {NoStop}%
\bibitem [{\citenamefont {Gnielinski}(1975)}]{gnielinski1975neue}%
  \BibitemOpen
  \bibfield  {author} {\bibinfo {author} {\bibfnamefont {V.}~\bibnamefont
  {Gnielinski}},\ }\bibfield  {title} {\enquote {\bibinfo {title} {New
  equations for heat and material transfer in turbulent flow pipes and
  channels},}\ }\href@noop {} {\bibfield  {journal} {\bibinfo  {journal}
  {Research in Engineering A}\ }\textbf {\bibinfo {volume} {41}},\ \bibinfo
  {pages} {8--16} (\bibinfo {year} {1975})}\BibitemShut {NoStop}%
\bibitem [{\citenamefont {Nunner}(1958)}]{nunner1958heat}%
  \BibitemOpen
  \bibfield  {author} {\bibinfo {author} {\bibfnamefont {W.}~\bibnamefont
  {Nunner}},\ }\href@noop {} {\emph {\bibinfo {title} {Heat transfer and
  pressure drop in rough tubes}}}\ (\bibinfo  {publisher} {Atomic Energy
  Research Establishment},\ \bibinfo {year} {1958})\BibitemShut {NoStop}%
\bibitem [{\citenamefont {Dipprey}\ and\ \citenamefont
  {Sabersky}(1963)}]{DIPPREY1963329}%
  \BibitemOpen
  \bibfield  {author} {\bibinfo {author} {\bibfnamefont {D.}~\bibnamefont
  {Dipprey}}\ and\ \bibinfo {author} {\bibfnamefont {R.}~\bibnamefont
  {Sabersky}},\ }\bibfield  {title} {\enquote {\bibinfo {title} {Heat and
  momentum transfer in smooth and rough tubes at various prandtl numbers},}\
  }\href {\doibase https://doi.org/10.1016/0017-9310(63)90097-8} {\bibfield
  {journal} {\bibinfo  {journal} {Int. J. Heat Mass Transf.}\ }\textbf
  {\bibinfo {volume} {6}},\ \bibinfo {pages} {329--353} (\bibinfo {year}
  {1963})}\BibitemShut {NoStop}%
\bibitem [{\citenamefont {Kays}, \citenamefont {Crawford},\ and\ \citenamefont
  {Weigand}(1980)}]{kays1980convective}%
  \BibitemOpen
  \bibfield  {author} {\bibinfo {author} {\bibfnamefont {W.~M.}\ \bibnamefont
  {Kays}}, \bibinfo {author} {\bibfnamefont {M.~E.}\ \bibnamefont {Crawford}},
  \ and\ \bibinfo {author} {\bibfnamefont {B.}~\bibnamefont {Weigand}},\
  }\href@noop {} {\emph {\bibinfo {title} {Convective heat and mass
  transfer}}},\ Vol.~\bibinfo {volume} {4}\ (\bibinfo  {publisher} {McGraw-Hill
  New York},\ \bibinfo {year} {1980})\BibitemShut {NoStop}%
\bibitem [{\citenamefont {Kays}, \citenamefont {Crawford},\ and\ \citenamefont
  {Weigand}(2005)}]{kays2005convective}%
  \BibitemOpen
  \bibfield  {author} {\bibinfo {author} {\bibfnamefont {W.~M.}\ \bibnamefont
  {Kays}}, \bibinfo {author} {\bibfnamefont {M.~E.}\ \bibnamefont {Crawford}},
  \ and\ \bibinfo {author} {\bibfnamefont {B.}~\bibnamefont {Weigand}},\
  }\href@noop {} {\emph {\bibinfo {title} {Convective heat and mass
  transfer}}},\ Vol.~\bibinfo {volume} {4}\ (\bibinfo  {publisher} {McGraw-Hill
  New York},\ \bibinfo {year} {2005})\BibitemShut {NoStop}%
\bibitem [{\citenamefont {Bons}(2002)}]{bons2002st}%
  \BibitemOpen
  \bibfield  {author} {\bibinfo {author} {\bibfnamefont {J.~P.}\ \bibnamefont
  {Bons}},\ }\bibfield  {title} {\enquote {\bibinfo {title} {$st$ and $c_f$
  augmentation for real turbine roughness with elevated freestream
  turbulence},}\ }in\ \href@noop {} {\emph {\bibinfo {booktitle} {ASME Turbo
  Expo.}}},\ Vol.\ \bibinfo {volume} {36088}\ (\bibinfo {year} {2002})\ pp.\
  \bibinfo {pages} {349--363}\BibitemShut {NoStop}%
\bibitem [{\citenamefont {Li}\ \emph {et~al.}(2007)\citenamefont {Li},
  \citenamefont {He}, \citenamefont {Tang},\ and\ \citenamefont
  {Tao}}]{li2007experimental}%
  \BibitemOpen
  \bibfield  {author} {\bibinfo {author} {\bibfnamefont {Z.}~\bibnamefont
  {Li}}, \bibinfo {author} {\bibfnamefont {Y.-L.}\ \bibnamefont {He}}, \bibinfo
  {author} {\bibfnamefont {G.-H.}\ \bibnamefont {Tang}}, \ and\ \bibinfo
  {author} {\bibfnamefont {W.-Q.}\ \bibnamefont {Tao}},\ }\bibfield  {title}
  {\enquote {\bibinfo {title} {Experimental and numerical studies of liquid
  flow and heat transfer in microtubes},}\ }\href@noop {} {\bibfield  {journal}
  {\bibinfo  {journal} {Int. J Heat Mass Transf.}\ }\textbf {\bibinfo {volume}
  {50}},\ \bibinfo {pages} {3447--3460} (\bibinfo {year} {2007})}\BibitemShut
  {NoStop}%
\bibitem [{\citenamefont {Pelevi{\'c}}\ and\ \citenamefont {van~der
  Meer}(2016)}]{pelevic2016heat}%
  \BibitemOpen
  \bibfield  {author} {\bibinfo {author} {\bibfnamefont {N.}~\bibnamefont
  {Pelevi{\'c}}}\ and\ \bibinfo {author} {\bibfnamefont {T.~H.}\ \bibnamefont
  {van~der Meer}},\ }\bibfield  {title} {\enquote {\bibinfo {title} {Heat
  transfer and pressure drop in microchannels with random roughness},}\
  }\href@noop {} {\bibfield  {journal} {\bibinfo  {journal} {Int. J Thermal
  Sci.}\ }\textbf {\bibinfo {volume} {99}},\ \bibinfo {pages} {125--135}
  (\bibinfo {year} {2016})}\BibitemShut {NoStop}%
\bibitem [{\citenamefont {Lu}\ \emph {et~al.}(2020)\citenamefont {Lu},
  \citenamefont {Xu}, \citenamefont {Gong}, \citenamefont {Duan},\ and\
  \citenamefont {Chai}}]{lu2020effects}%
  \BibitemOpen
  \bibfield  {author} {\bibinfo {author} {\bibfnamefont {H.}~\bibnamefont
  {Lu}}, \bibinfo {author} {\bibfnamefont {M.}~\bibnamefont {Xu}}, \bibinfo
  {author} {\bibfnamefont {L.}~\bibnamefont {Gong}}, \bibinfo {author}
  {\bibfnamefont {X.}~\bibnamefont {Duan}}, \ and\ \bibinfo {author}
  {\bibfnamefont {J.~C.}\ \bibnamefont {Chai}},\ }\bibfield  {title} {\enquote
  {\bibinfo {title} {Effects of surface roughness in microchannel with passive
  heat transfer enhancement structures},}\ }\href@noop {} {\bibfield  {journal}
  {\bibinfo  {journal} {International Journal of Heat and Mass Transfer}\
  }\textbf {\bibinfo {volume} {148}},\ \bibinfo {pages} {119070} (\bibinfo
  {year} {2020})}\BibitemShut {NoStop}%
\bibitem [{\citenamefont {Ansari}\ and\ \citenamefont
  {Zhou}(2020)}]{ansari2020influence}%
  \BibitemOpen
  \bibfield  {author} {\bibinfo {author} {\bibfnamefont {M.~Q.}\ \bibnamefont
  {Ansari}}\ and\ \bibinfo {author} {\bibfnamefont {G.}~\bibnamefont {Zhou}},\
  }\bibfield  {title} {\enquote {\bibinfo {title} {Influence of structured
  surface roughness peaks on flow and heat transfer performances of micro-and
  mini-channels},}\ }\href@noop {} {\bibfield  {journal} {\bibinfo  {journal}
  {Int. Commun. Heat Mass Transf.}\ }\textbf {\bibinfo {volume} {110}},\
  \bibinfo {pages} {104428} (\bibinfo {year} {2020})}\BibitemShut {NoStop}%
\bibitem [{\citenamefont {Croce}, \citenamefont {D’agaro},\ and\
  \citenamefont {Nonino}(2007)}]{croce2007three}%
  \BibitemOpen
  \bibfield  {author} {\bibinfo {author} {\bibfnamefont {G.}~\bibnamefont
  {Croce}}, \bibinfo {author} {\bibfnamefont {P.}~\bibnamefont {D’agaro}}, \
  and\ \bibinfo {author} {\bibfnamefont {C.}~\bibnamefont {Nonino}},\
  }\bibfield  {title} {\enquote {\bibinfo {title} {Three-dimensional roughness
  effect on microchannel heat transfer and pressure drop},}\ }\href@noop {}
  {\bibfield  {journal} {\bibinfo  {journal} {Int. J Heat Mass Transf.}\
  }\textbf {\bibinfo {volume} {50}},\ \bibinfo {pages} {5249--5259} (\bibinfo
  {year} {2007})}\BibitemShut {NoStop}%
\bibitem [{\citenamefont {Xiong}\ and\ \citenamefont
  {Chung}(2010)}]{xiong2010investigation}%
  \BibitemOpen
  \bibfield  {author} {\bibinfo {author} {\bibfnamefont {R.}~\bibnamefont
  {Xiong}}\ and\ \bibinfo {author} {\bibfnamefont {J.~N.}\ \bibnamefont
  {Chung}},\ }\bibfield  {title} {\enquote {\bibinfo {title} {Investigation of
  laminar flow in microtubes with random rough surfaces},}\ }\href@noop {}
  {\bibfield  {journal} {\bibinfo  {journal} {Microfluidics and Nanofluidics}\
  }\textbf {\bibinfo {volume} {8}},\ \bibinfo {pages} {11--20} (\bibinfo {year}
  {2010})}\BibitemShut {NoStop}%
\bibitem [{\citenamefont {Kadivar}, \citenamefont {Tormey},\ and\ \citenamefont
  {McGranaghan}(2022)}]{kadivar2022cfd}%
  \BibitemOpen
  \bibfield  {author} {\bibinfo {author} {\bibfnamefont {M.}~\bibnamefont
  {Kadivar}}, \bibinfo {author} {\bibfnamefont {D.}~\bibnamefont {Tormey}}, \
  and\ \bibinfo {author} {\bibfnamefont {G.}~\bibnamefont {McGranaghan}},\
  }\bibfield  {title} {\enquote {\bibinfo {title} {Cfd of roughness effects on
  laminar heat transfer applied to additive manufactured minichannels},}\
  }\href@noop {} {\bibfield  {journal} {\bibinfo  {journal} {Heat Mass
  Transf.}\ ,\ \bibinfo {pages} {1--15}} (\bibinfo {year} {2022})}\BibitemShut
  {NoStop}%
\bibitem [{\citenamefont {Kadivar}, \citenamefont {Tormey},\ and\ \citenamefont
  {McGranaghan}(2023)}]{kadivar2023comparison}%
  \BibitemOpen
  \bibfield  {author} {\bibinfo {author} {\bibfnamefont {M.}~\bibnamefont
  {Kadivar}}, \bibinfo {author} {\bibfnamefont {D.}~\bibnamefont {Tormey}}, \
  and\ \bibinfo {author} {\bibfnamefont {G.}~\bibnamefont {McGranaghan}},\
  }\bibfield  {title} {\enquote {\bibinfo {title} {A comparison of rans models
  used for cfd prediction of turbulent flow and heat transfer in rough and
  smooth channels},}\ }\href@noop {} {\bibfield  {journal} {\bibinfo  {journal}
  {International Journal of Thermofluids}\ }\textbf {\bibinfo {volume} {20}},\
  \bibinfo {pages} {100399} (\bibinfo {year} {2023})}\BibitemShut {NoStop}%
\bibitem [{\citenamefont {Garg}, \citenamefont {Wang},\ and\ \citenamefont
  {Fureby}(2023)}]{garg2023largeribs}%
  \BibitemOpen
  \bibfield  {author} {\bibinfo {author} {\bibfnamefont {H.}~\bibnamefont
  {Garg}}, \bibinfo {author} {\bibfnamefont {L.}~\bibnamefont {Wang}}, \ and\
  \bibinfo {author} {\bibfnamefont {C.}~\bibnamefont {Fureby}},\ }\bibfield
  {title} {\enquote {\bibinfo {title} {Large-eddy simulations of separated flow
  and heat transfer in a ribbed channel},}\ }in\ \href@noop {} {\emph {\bibinfo
  {booktitle} {ICHMT Digital Library Online}}}\ (\bibinfo {organization} {Begel
  House Inc.},\ \bibinfo {year} {2023})\BibitemShut {NoStop}%
\bibitem [{\citenamefont {Forooghi}\ \emph {et~al.}(2018)\citenamefont
  {Forooghi}, \citenamefont {Stroh}, \citenamefont {Schlatter},\ and\
  \citenamefont {Frohnapfel}}]{forooghi2018direct}%
  \BibitemOpen
  \bibfield  {author} {\bibinfo {author} {\bibfnamefont {P.}~\bibnamefont
  {Forooghi}}, \bibinfo {author} {\bibfnamefont {A.}~\bibnamefont {Stroh}},
  \bibinfo {author} {\bibfnamefont {P.}~\bibnamefont {Schlatter}}, \ and\
  \bibinfo {author} {\bibfnamefont {B.}~\bibnamefont {Frohnapfel}},\ }\bibfield
   {title} {\enquote {\bibinfo {title} {Direct numerical simulation of flow
  over dissimilar, randomly distributed roughness elements: A systematic study
  on the effect of surface morphology on turbulence},}\ }\href@noop {}
  {\bibfield  {journal} {\bibinfo  {journal} {Phys. Rev. Fluids}\ }\textbf
  {\bibinfo {volume} {3}},\ \bibinfo {pages} {044605} (\bibinfo {year}
  {2018})}\BibitemShut {NoStop}%
\bibitem [{\citenamefont {Peeters}\ and\ \citenamefont
  {Sandham}(2019)}]{peeters2019turbulent}%
  \BibitemOpen
  \bibfield  {author} {\bibinfo {author} {\bibfnamefont {J.~W.~R.}\
  \bibnamefont {Peeters}}\ and\ \bibinfo {author} {\bibfnamefont {N.~D.}\
  \bibnamefont {Sandham}},\ }\bibfield  {title} {\enquote {\bibinfo {title}
  {Turbulent heat transfer in channels with irregular roughness},}\ }\href@noop
  {} {\bibfield  {journal} {\bibinfo  {journal} {Int. J. Heat Mass Transf.}\
  }\textbf {\bibinfo {volume} {138}},\ \bibinfo {pages} {454--467} (\bibinfo
  {year} {2019})}\BibitemShut {NoStop}%
\bibitem [{\citenamefont {MacDonald}\ \emph {et~al.}(2019)\citenamefont
  {MacDonald}, \citenamefont {Hutchins}, \citenamefont {Lohse},\ and\
  \citenamefont {Chung}}]{macdonald2019heat}%
  \BibitemOpen
  \bibfield  {author} {\bibinfo {author} {\bibfnamefont {M.}~\bibnamefont
  {MacDonald}}, \bibinfo {author} {\bibfnamefont {N.}~\bibnamefont {Hutchins}},
  \bibinfo {author} {\bibfnamefont {D.}~\bibnamefont {Lohse}}, \ and\ \bibinfo
  {author} {\bibfnamefont {D.}~\bibnamefont {Chung}},\ }\bibfield  {title}
  {\enquote {\bibinfo {title} {Heat transfer in rough-wall turbulent thermal
  convection in the ultimate regime},}\ }\href@noop {} {\bibfield  {journal}
  {\bibinfo  {journal} {Phys. Rev. Fluids}\ }\textbf {\bibinfo {volume} {4}},\
  \bibinfo {pages} {071501} (\bibinfo {year} {2019})}\BibitemShut {NoStop}%
\bibitem [{\citenamefont {Kuwata}(2021)}]{kuwata2021direct}%
  \BibitemOpen
  \bibfield  {author} {\bibinfo {author} {\bibfnamefont {Y.}~\bibnamefont
  {Kuwata}},\ }\bibfield  {title} {\enquote {\bibinfo {title} {Direct numerical
  simulation of turbulent heat transfer on the reynolds analogy over irregular
  rough surfaces},}\ }\href@noop {} {\bibfield  {journal} {\bibinfo  {journal}
  {Int. J. Heat Fluid Flow}\ }\textbf {\bibinfo {volume} {92}},\ \bibinfo
  {pages} {108859} (\bibinfo {year} {2021})}\BibitemShut {NoStop}%
\bibitem [{\citenamefont {Garg}, \citenamefont {Wang},\ and\ \citenamefont
  {Fureby}(2024)}]{garg2024heat}%
  \BibitemOpen
  \bibfield  {author} {\bibinfo {author} {\bibfnamefont {H.}~\bibnamefont
  {Garg}}, \bibinfo {author} {\bibfnamefont {L.}~\bibnamefont {Wang}}, \ and\
  \bibinfo {author} {\bibfnamefont {C.}~\bibnamefont {Fureby}},\ }\bibfield
  {title} {\enquote {\bibinfo {title} {Heat transfer enhancement with
  additively manufactured rough surfaces: Insights from large-eddy
  simulations},}\ }\href@noop {} {\bibfield  {journal} {\bibinfo  {journal}
  {Phys. Fluids}\ }\textbf {\bibinfo {volume} {36}} (\bibinfo {year}
  {2024})}\BibitemShut {NoStop}%
\bibitem [{\citenamefont {Stimpson}\ \emph {et~al.}(2017)\citenamefont
  {Stimpson}, \citenamefont {Snyder}, \citenamefont {Thole},\ and\
  \citenamefont {Mongillo}}]{stimpson2017scaling}%
  \BibitemOpen
  \bibfield  {author} {\bibinfo {author} {\bibfnamefont {C.~K.}\ \bibnamefont
  {Stimpson}}, \bibinfo {author} {\bibfnamefont {J.~C.}\ \bibnamefont
  {Snyder}}, \bibinfo {author} {\bibfnamefont {K.~A.}\ \bibnamefont {Thole}}, \
  and\ \bibinfo {author} {\bibfnamefont {D.}~\bibnamefont {Mongillo}},\
  }\bibfield  {title} {\enquote {\bibinfo {title} {{Scaling roughness effects
  on pressure loss and heat transfer of additively manufactured channels}},}\
  }\href@noop {} {\bibfield  {journal} {\bibinfo  {journal} {J. Turbomach.}\
  }\textbf {\bibinfo {volume} {139}} (\bibinfo {year} {2017})}\BibitemShut
  {NoStop}%
\bibitem [{\citenamefont {Snyder}\ and\ \citenamefont
  {Thole}(2020)}]{snyder2020tailoring}%
  \BibitemOpen
  \bibfield  {author} {\bibinfo {author} {\bibfnamefont {J.~C.}\ \bibnamefont
  {Snyder}}\ and\ \bibinfo {author} {\bibfnamefont {K.~A.}\ \bibnamefont
  {Thole}},\ }\bibfield  {title} {\enquote {\bibinfo {title} {{Tailoring
  surface roughness using additive manufacturing to improve internal
  cooling}},}\ }\href@noop {} {\bibfield  {journal} {\bibinfo  {journal} {J.
  Turbomach.}\ }\textbf {\bibinfo {volume} {142}} (\bibinfo {year}
  {2020})}\BibitemShut {NoStop}%
\bibitem [{\citenamefont {McClain}\ \emph {et~al.}(2021)\citenamefont
  {McClain}, \citenamefont {Hanson}, \citenamefont {Cinnamon}, \citenamefont
  {Snyder}, \citenamefont {Kunz},\ and\ \citenamefont
  {Thole}}]{mcclain2021flow}%
  \BibitemOpen
  \bibfield  {author} {\bibinfo {author} {\bibfnamefont {S.~T.}\ \bibnamefont
  {McClain}}, \bibinfo {author} {\bibfnamefont {D.~R.}\ \bibnamefont {Hanson}},
  \bibinfo {author} {\bibfnamefont {E.}~\bibnamefont {Cinnamon}}, \bibinfo
  {author} {\bibfnamefont {J.~C.}\ \bibnamefont {Snyder}}, \bibinfo {author}
  {\bibfnamefont {R.~F.}\ \bibnamefont {Kunz}}, \ and\ \bibinfo {author}
  {\bibfnamefont {K.~A.}\ \bibnamefont {Thole}},\ }\bibfield  {title} {\enquote
  {\bibinfo {title} {{Flow in a simulated turbine blade cooling channel with
  spatially varying roughness caused by additive manufacturing orientation}},}\
  }\href@noop {} {\bibfield  {journal} {\bibinfo  {journal} {J. Turbomach.}\
  }\textbf {\bibinfo {volume} {143}} (\bibinfo {year} {2021})}\BibitemShut
  {NoStop}%
\bibitem [{\citenamefont {Favero}\ \emph {et~al.}(2022)\citenamefont {Favero},
  \citenamefont {Berti}, \citenamefont {Bonesso}, \citenamefont {Morrone},
  \citenamefont {Oriolo}, \citenamefont {Rebesan}, \citenamefont {Dima},
  \citenamefont {Gregori}, \citenamefont {Pepato}, \citenamefont {Scanavini},\
  and\ \citenamefont {Mancin}}]{FAVERO2022106128}%
  \BibitemOpen
  \bibfield  {author} {\bibinfo {author} {\bibfnamefont {G.}~\bibnamefont
  {Favero}}, \bibinfo {author} {\bibfnamefont {G.}~\bibnamefont {Berti}},
  \bibinfo {author} {\bibfnamefont {M.}~\bibnamefont {Bonesso}}, \bibinfo
  {author} {\bibfnamefont {D.}~\bibnamefont {Morrone}}, \bibinfo {author}
  {\bibfnamefont {S.}~\bibnamefont {Oriolo}}, \bibinfo {author} {\bibfnamefont
  {P.}~\bibnamefont {Rebesan}}, \bibinfo {author} {\bibfnamefont
  {R.}~\bibnamefont {Dima}}, \bibinfo {author} {\bibfnamefont {P.}~\bibnamefont
  {Gregori}}, \bibinfo {author} {\bibfnamefont {A.}~\bibnamefont {Pepato}},
  \bibinfo {author} {\bibfnamefont {A.}~\bibnamefont {Scanavini}}, \ and\
  \bibinfo {author} {\bibfnamefont {S.}~\bibnamefont {Mancin}},\ }\bibfield
  {title} {\enquote {\bibinfo {title} {{Experimental and numerical analyses of
  fluid flow inside additively manufactured and smoothed cooling channels}},}\
  }\href@noop {} {\bibfield  {journal} {\bibinfo  {journal} {Int. Commun. Heat
  Mass Trans.}\ }\textbf {\bibinfo {volume} {135}},\ \bibinfo {pages} {106128}
  (\bibinfo {year} {2022})}\BibitemShut {NoStop}%
\bibitem [{\citenamefont {Garg}\ \emph {et~al.}(2023)\citenamefont {Garg},
  \citenamefont {Wang}, \citenamefont {Sahut},\ and\ \citenamefont
  {Fureby}}]{garg2023large}%
  \BibitemOpen
  \bibfield  {author} {\bibinfo {author} {\bibfnamefont {H.}~\bibnamefont
  {Garg}}, \bibinfo {author} {\bibfnamefont {L.}~\bibnamefont {Wang}}, \bibinfo
  {author} {\bibfnamefont {G.}~\bibnamefont {Sahut}}, \ and\ \bibinfo {author}
  {\bibfnamefont {C.}~\bibnamefont {Fureby}},\ }\bibfield  {title} {\enquote
  {\bibinfo {title} {Large eddy simulations of fully developed turbulent flows
  over additively manufactured rough surfaces},}\ }\href@noop {} {\bibfield
  {journal} {\bibinfo  {journal} {Physics of Fluids}\ }\textbf {\bibinfo
  {volume} {35}} (\bibinfo {year} {2023})}\BibitemShut {NoStop}%
\bibitem [{Note1()}]{Note1}%
  \BibitemOpen
  \bibinfo {note} {\unhbox \voidb@x \begingroup \begingroup \let \relax \relax
  \relax \endgroup \protect \Url
  {https://github.com/CoffeeDynamics/STLRoughPipes}}\BibitemShut {NoStop}%
\bibitem [{\citenamefont {Nicoud}\ and\ \citenamefont
  {Ducros}(1999)}]{nicoud1999subgrid}%
  \BibitemOpen
  \bibfield  {author} {\bibinfo {author} {\bibfnamefont {F.}~\bibnamefont
  {Nicoud}}\ and\ \bibinfo {author} {\bibfnamefont {F.}~\bibnamefont
  {Ducros}},\ }\bibfield  {title} {\enquote {\bibinfo {title} {Subgrid-scale
  stress modelling based on the square of the velocity gradient tensor},}\
  }\href@noop {} {\bibfield  {journal} {\bibinfo  {journal} {Flow Turbul.
  Combust.}\ }\textbf {\bibinfo {volume} {62}},\ \bibinfo {pages} {183--200}
  (\bibinfo {year} {1999})}\BibitemShut {NoStop}%
\bibitem [{\citenamefont {Moin}\ \emph {et~al.}(1991)\citenamefont {Moin},
  \citenamefont {Squires}, \citenamefont {Cabot},\ and\ \citenamefont
  {Lee}}]{moin1991dynamic}%
  \BibitemOpen
  \bibfield  {author} {\bibinfo {author} {\bibfnamefont {P.}~\bibnamefont
  {Moin}}, \bibinfo {author} {\bibfnamefont {K.}~\bibnamefont {Squires}},
  \bibinfo {author} {\bibfnamefont {W.}~\bibnamefont {Cabot}}, \ and\ \bibinfo
  {author} {\bibfnamefont {S.}~\bibnamefont {Lee}},\ }\bibfield  {title}
  {\enquote {\bibinfo {title} {A dynamic subgrid-scale model for compressible
  turbulence and scalar transport},}\ }\href@noop {} {\bibfield  {journal}
  {\bibinfo  {journal} {Phys. Fluids A}\ }\textbf {\bibinfo {volume} {3}},\
  \bibinfo {pages} {2746--2757} (\bibinfo {year} {1991})}\BibitemShut {NoStop}%
\bibitem [{\citenamefont {{Kasagi}}, \citenamefont {{Tomita}},\ and\
  \citenamefont {{Kuroda}}(1992)}]{kasagi1992}%
  \BibitemOpen
  \bibfield  {author} {\bibinfo {author} {\bibfnamefont {N.}~\bibnamefont
  {{Kasagi}}}, \bibinfo {author} {\bibfnamefont {Y.}~\bibnamefont {{Tomita}}},
  \ and\ \bibinfo {author} {\bibfnamefont {A.}~\bibnamefont {{Kuroda}}},\
  }\bibfield  {title} {\enquote {\bibinfo {title} {{Direct numerical simulation
  of passive scalar field in a turbulent channel flow}},}\ }\href@noop {}
  {\bibfield  {journal} {\bibinfo  {journal} {ASME J. Heat Trans.}\ }\textbf
  {\bibinfo {volume} {114}},\ \bibinfo {pages} {598--606} (\bibinfo {year}
  {1992})}\BibitemShut {NoStop}%
\bibitem [{\citenamefont {Kozuka}, \citenamefont {Seki},\ and\ \citenamefont
  {Kawamura}(2009)}]{kozuka2009}%
  \BibitemOpen
  \bibfield  {author} {\bibinfo {author} {\bibfnamefont {M.}~\bibnamefont
  {Kozuka}}, \bibinfo {author} {\bibfnamefont {Y.}~\bibnamefont {Seki}}, \ and\
  \bibinfo {author} {\bibfnamefont {H.}~\bibnamefont {Kawamura}},\ }\bibfield
  {title} {\enquote {\bibinfo {title} {{DNS of turbulent heat transfer in a
  channel flow with a high spatial resolution}},}\ }\href@noop {} {\bibfield
  {journal} {\bibinfo  {journal} {Int. J. Heat Fluid Flow}\ }\textbf {\bibinfo
  {volume} {30}},\ \bibinfo {pages} {514--524} (\bibinfo {year}
  {2009})}\BibitemShut {NoStop}%
\bibitem [{\citenamefont {Lluesma-Rodr{\'\i}guez}, \citenamefont {Hoyas},\ and\
  \citenamefont {Perez-Quiles}(2018)}]{lluesma2018}%
  \BibitemOpen
  \bibfield  {author} {\bibinfo {author} {\bibfnamefont {F.}~\bibnamefont
  {Lluesma-Rodr{\'\i}guez}}, \bibinfo {author} {\bibfnamefont {S.}~\bibnamefont
  {Hoyas}}, \ and\ \bibinfo {author} {\bibfnamefont {M.~J.}\ \bibnamefont
  {Perez-Quiles}},\ }\bibfield  {title} {\enquote {\bibinfo {title} {{Influence
  of the computational domain on DNS of turbulent heat transfer up to
  Re$_\tau$= 2000 for Pr= 0.71}},}\ }\href@noop {} {\bibfield  {journal}
  {\bibinfo  {journal} {Int. J. Heat Mass Transf.}\ }\textbf {\bibinfo {volume}
  {122}},\ \bibinfo {pages} {983--992} (\bibinfo {year} {2018})}\BibitemShut
  {NoStop}%
\bibitem [{\citenamefont {Butcher}(2016)}]{butcher2016numerical}%
  \BibitemOpen
  \bibfield  {author} {\bibinfo {author} {\bibfnamefont {J.~C.}\ \bibnamefont
  {Butcher}},\ }\href@noop {} {\emph {\bibinfo {title} {{Numerical methods for
  ordinary differential equations}}}}\ (\bibinfo  {publisher} {John Wiley \&
  Sons},\ \bibinfo {year} {2016})\BibitemShut {NoStop}%
\bibitem [{\citenamefont {Weller}\ \emph {et~al.}(1998)\citenamefont {Weller},
  \citenamefont {Tabor}, \citenamefont {Jasak},\ and\ \citenamefont
  {Fureby}}]{weller1998tensorial}%
  \BibitemOpen
  \bibfield  {author} {\bibinfo {author} {\bibfnamefont {H.~G.}\ \bibnamefont
  {Weller}}, \bibinfo {author} {\bibfnamefont {G.}~\bibnamefont {Tabor}},
  \bibinfo {author} {\bibfnamefont {H.}~\bibnamefont {Jasak}}, \ and\ \bibinfo
  {author} {\bibfnamefont {C.}~\bibnamefont {Fureby}},\ }\bibfield  {title}
  {\enquote {\bibinfo {title} {{A tensorial approach to computational continuum
  mechanics using object-oriented techniques}},}\ }\href@noop {} {\bibfield
  {journal} {\bibinfo  {journal} {Comp. Phys.}\ }\textbf {\bibinfo {volume}
  {12}},\ \bibinfo {pages} {620--631} (\bibinfo {year} {1998})}\BibitemShut
  {NoStop}%
\bibitem [{\citenamefont {Ruge}\ and\ \citenamefont {Stüben}(1987)}]{GAMG}%
  \BibitemOpen
  \bibfield  {author} {\bibinfo {author} {\bibfnamefont {J.~W.}\ \bibnamefont
  {Ruge}}\ and\ \bibinfo {author} {\bibfnamefont {K.}~\bibnamefont {Stüben}},\
  }\enquote {\bibinfo {title} {4. algebraic multigrid},}\ in\ \href@noop {}
  {\emph {\bibinfo {booktitle} {Multigrid Methods}}}\ (\bibinfo {year} {1987})\
  pp.\ \bibinfo {pages} {73--130}\BibitemShut {NoStop}%
\bibitem [{\citenamefont {Golub}\ and\ \citenamefont
  {Van~Loan}(1996)}]{van1996matrix}%
  \BibitemOpen
  \bibfield  {author} {\bibinfo {author} {\bibfnamefont {G.~H.}\ \bibnamefont
  {Golub}}\ and\ \bibinfo {author} {\bibfnamefont {C.~F.}\ \bibnamefont
  {Van~Loan}},\ }\bibfield  {title} {\enquote {\bibinfo {title} {{Matrix
  computations (Johns Hopkins studies in mathematical sciences)}},}\
  }\href@noop {} {\bibfield  {journal} {\bibinfo  {journal} {Matrix Comput.}\ }
  (\bibinfo {year} {1996})}\BibitemShut {NoStop}%
\bibitem [{\citenamefont {Press}\ \emph {et~al.}(2007)\citenamefont {Press},
  \citenamefont {Teukolsky}, \citenamefont {Vetterling},\ and\ \citenamefont
  {Flannery}}]{press2007numerical}%
  \BibitemOpen
  \bibfield  {author} {\bibinfo {author} {\bibfnamefont {W.~H.}\ \bibnamefont
  {Press}}, \bibinfo {author} {\bibfnamefont {S.~A.}\ \bibnamefont
  {Teukolsky}}, \bibinfo {author} {\bibfnamefont {W.~T.}\ \bibnamefont
  {Vetterling}}, \ and\ \bibinfo {author} {\bibfnamefont {B.~P.}\ \bibnamefont
  {Flannery}},\ }\href@noop {} {\emph {\bibinfo {title} {{Numerical recipes 3rd
  edition: The art of scientific computing}}}}\ (\bibinfo  {publisher}
  {Cambridge university press},\ \bibinfo {year} {2007})\BibitemShut {NoStop}%
\bibitem [{\citenamefont {Lozano}\ and\ \citenamefont
  {Jim{\'e}nez}(2014)}]{lozano2014effect}%
  \BibitemOpen
  \bibfield  {author} {\bibinfo {author} {\bibfnamefont {D.~A.}\ \bibnamefont
  {Lozano}}\ and\ \bibinfo {author} {\bibfnamefont {J.}~\bibnamefont
  {Jim{\'e}nez}},\ }\bibfield  {title} {\enquote {\bibinfo {title} {Effect of
  the computational domain on direct simulations of turbulent channels up to re
  $\tau$= 4200},}\ }\href@noop {} {\bibfield  {journal} {\bibinfo  {journal}
  {Phys. Fluids}\ }\textbf {\bibinfo {volume} {26}},\ \bibinfo {pages} {011702}
  (\bibinfo {year} {2014})}\BibitemShut {NoStop}%
\bibitem [{\citenamefont {Lluesma}, \citenamefont {Hoyas},\ and\ \citenamefont
  {Perez}(2018)}]{lluesma2018influence}%
  \BibitemOpen
  \bibfield  {author} {\bibinfo {author} {\bibfnamefont {R.~F.}\ \bibnamefont
  {Lluesma}}, \bibinfo {author} {\bibfnamefont {S.}~\bibnamefont {Hoyas}}, \
  and\ \bibinfo {author} {\bibfnamefont {M.~J.~Q.}\ \bibnamefont {Perez}},\
  }\bibfield  {title} {\enquote {\bibinfo {title} {Influence of the
  computational domain on dns of turbulent heat transfer up to re$\tau$= 2000
  for pr= 0.71},}\ }\href@noop {} {\bibfield  {journal} {\bibinfo  {journal}
  {Int. J. Heat Mass Transf.}\ }\textbf {\bibinfo {volume} {122}},\ \bibinfo
  {pages} {983--992} (\bibinfo {year} {2018})}\BibitemShut {NoStop}%
\bibitem [{\citenamefont {Garg}\ \emph {et~al.}(2024)\citenamefont {Garg},
  \citenamefont {Wang}, \citenamefont {Andersson},\ and\ \citenamefont
  {Fureby}}]{garg2024large}%
  \BibitemOpen
  \bibfield  {author} {\bibinfo {author} {\bibfnamefont {H.}~\bibnamefont
  {Garg}}, \bibinfo {author} {\bibfnamefont {L.}~\bibnamefont {Wang}}, \bibinfo
  {author} {\bibfnamefont {M.}~\bibnamefont {Andersson}}, \ and\ \bibinfo
  {author} {\bibfnamefont {C.}~\bibnamefont {Fureby}},\ }\bibfield  {title}
  {\enquote {\bibinfo {title} {Large eddy simulations of turbulent pipe flows
  at moderate reynolds numbers},}\ }\href@noop {} {\bibfield  {journal}
  {\bibinfo  {journal} {Physics of Fluids}\ }\textbf {\bibinfo {volume} {36}}
  (\bibinfo {year} {2024})}\BibitemShut {NoStop}%
\bibitem [{\citenamefont {Raupach}(1994)}]{raupach1994simplified}%
  \BibitemOpen
  \bibfield  {author} {\bibinfo {author} {\bibfnamefont {M.~R.}\ \bibnamefont
  {Raupach}},\ }\bibfield  {title} {\enquote {\bibinfo {title} {Simplified
  expressions for vegetation roughness length and zero-plane displacement as
  functions of canopy height and area index},}\ }\href@noop {} {\bibfield
  {journal} {\bibinfo  {journal} {Boundary Layer Meteorol.}\ }\textbf {\bibinfo
  {volume} {71}},\ \bibinfo {pages} {211--216} (\bibinfo {year}
  {1994})}\BibitemShut {NoStop}%
\bibitem [{\citenamefont {MacDonald}, \citenamefont {Hutchins},\ and\
  \citenamefont {Chung}(2019)}]{macdonald2019roughness}%
  \BibitemOpen
  \bibfield  {author} {\bibinfo {author} {\bibfnamefont {M.}~\bibnamefont
  {MacDonald}}, \bibinfo {author} {\bibfnamefont {N.}~\bibnamefont {Hutchins}},
  \ and\ \bibinfo {author} {\bibfnamefont {D.}~\bibnamefont {Chung}},\
  }\bibfield  {title} {\enquote {\bibinfo {title} {Roughness effects in
  turbulent forced convection},}\ }\href@noop {} {\bibfield  {journal}
  {\bibinfo  {journal} {J. Fluid Mech.}\ }\textbf {\bibinfo {volume} {861}},\
  \bibinfo {pages} {138--162} (\bibinfo {year} {2019})}\BibitemShut {NoStop}%
\bibitem [{\citenamefont {Forooghi}, \citenamefont {Stripf},\ and\
  \citenamefont {Frohnapfel}(2018)}]{forooghi2018systematic}%
  \BibitemOpen
  \bibfield  {author} {\bibinfo {author} {\bibfnamefont {P.}~\bibnamefont
  {Forooghi}}, \bibinfo {author} {\bibfnamefont {M.}~\bibnamefont {Stripf}}, \
  and\ \bibinfo {author} {\bibfnamefont {B.}~\bibnamefont {Frohnapfel}},\
  }\bibfield  {title} {\enquote {\bibinfo {title} {A systematic study of
  turbulent heat transfer over rough walls},}\ }\href@noop {} {\bibfield
  {journal} {\bibinfo  {journal} {Int. J. Heat Mass Transf.}\ }\textbf
  {\bibinfo {volume} {127}},\ \bibinfo {pages} {1157--1168} (\bibinfo {year}
  {2018})}\BibitemShut {NoStop}%
\bibitem [{\citenamefont {Busse}, \citenamefont {Thakkar},\ and\ \citenamefont
  {Sandham}(2017)}]{busse2017reynolds}%
  \BibitemOpen
  \bibfield  {author} {\bibinfo {author} {\bibfnamefont {A.}~\bibnamefont
  {Busse}}, \bibinfo {author} {\bibfnamefont {M.}~\bibnamefont {Thakkar}}, \
  and\ \bibinfo {author} {\bibfnamefont {N.~D.}\ \bibnamefont {Sandham}},\
  }\bibfield  {title} {\enquote {\bibinfo {title} {{Reynolds-number dependence
  of the near-wall flow over irregular rough surfaces}},}\ }\href@noop {}
  {\bibfield  {journal} {\bibinfo  {journal} {J. Fluid Mech.}\ }\textbf
  {\bibinfo {volume} {810}},\ \bibinfo {pages} {196--224} (\bibinfo {year}
  {2017})}\BibitemShut {NoStop}%
\bibitem [{\citenamefont {Thakkar}, \citenamefont {Busse},\ and\ \citenamefont
  {Sandham}(2017)}]{thakkar2017surface}%
  \BibitemOpen
  \bibfield  {author} {\bibinfo {author} {\bibfnamefont {M.}~\bibnamefont
  {Thakkar}}, \bibinfo {author} {\bibfnamefont {A.}~\bibnamefont {Busse}}, \
  and\ \bibinfo {author} {\bibfnamefont {N.}~\bibnamefont {Sandham}},\
  }\bibfield  {title} {\enquote {\bibinfo {title} {{Surface correlations of
  hydrodynamic drag for transitionally rough engineering surfaces}},}\
  }\href@noop {} {\bibfield  {journal} {\bibinfo  {journal} {J. Turb.}\
  }\textbf {\bibinfo {volume} {18}},\ \bibinfo {pages} {138--169} (\bibinfo
  {year} {2017})}\BibitemShut {NoStop}%
\bibitem [{\citenamefont {Barros}\ and\ \citenamefont
  {Christensen}(2019)}]{barros2019characteristics}%
  \BibitemOpen
  \bibfield  {author} {\bibinfo {author} {\bibfnamefont {J.~M.}\ \bibnamefont
  {Barros}}\ and\ \bibinfo {author} {\bibfnamefont {K.~T.}\ \bibnamefont
  {Christensen}},\ }\bibfield  {title} {\enquote {\bibinfo {title}
  {{Characteristics of large-scale and superstructure motions in a turbulent
  boundary layer overlying complex roughness}},}\ }\href@noop {} {\bibfield
  {journal} {\bibinfo  {journal} {J. Turb.}\ }\textbf {\bibinfo {volume}
  {20}},\ \bibinfo {pages} {147--173} (\bibinfo {year} {2019})}\BibitemShut
  {NoStop}%
\bibitem [{\citenamefont {Busse}\ and\ \citenamefont
  {Jelly}(2020)}]{busse2020influence}%
  \BibitemOpen
  \bibfield  {author} {\bibinfo {author} {\bibfnamefont {A.}~\bibnamefont
  {Busse}}\ and\ \bibinfo {author} {\bibfnamefont {T.~O.}\ \bibnamefont
  {Jelly}},\ }\bibfield  {title} {\enquote {\bibinfo {title} {Influence of
  surface anisotropy on turbulent flow over irregular roughness},}\ }\href@noop
  {} {\bibfield  {journal} {\bibinfo  {journal} {Flow Turb. Combust.}\ }\textbf
  {\bibinfo {volume} {104}},\ \bibinfo {pages} {331--354} (\bibinfo {year}
  {2020})}\BibitemShut {NoStop}%
\bibitem [{\citenamefont {Kuwata}(2022)}]{kuwata2022reynolds}%
  \BibitemOpen
  \bibfield  {author} {\bibinfo {author} {\bibfnamefont {Y.}~\bibnamefont
  {Kuwata}},\ }\bibfield  {title} {\enquote {\bibinfo {title} {Reynolds number
  dependence of turbulent heat transfer over irregular rough surfaces},}\
  }\href@noop {} {\bibfield  {journal} {\bibinfo  {journal} {Phys. Fluids}\
  }\textbf {\bibinfo {volume} {34}} (\bibinfo {year} {2022})}\BibitemShut
  {NoStop}%
\bibitem [{\citenamefont {Kader}(1981)}]{kader1981temperature}%
  \BibitemOpen
  \bibfield  {author} {\bibinfo {author} {\bibfnamefont {B.~A.}\ \bibnamefont
  {Kader}},\ }\bibfield  {title} {\enquote {\bibinfo {title} {Temperature and
  concentration profiles in fully turbulent boundary layers},}\ }\href@noop {}
  {\bibfield  {journal} {\bibinfo  {journal} {Int. J. Heat Mass Transf.}\
  }\textbf {\bibinfo {volume} {24}},\ \bibinfo {pages} {1541--1544} (\bibinfo
  {year} {1981})}\BibitemShut {NoStop}%
\bibitem [{\citenamefont {Spalart}\ and\ \citenamefont
  {Strelets}(2000)}]{spalart2000mechanisms}%
  \BibitemOpen
  \bibfield  {author} {\bibinfo {author} {\bibfnamefont {P.~R.}\ \bibnamefont
  {Spalart}}\ and\ \bibinfo {author} {\bibfnamefont {M.~K.}\ \bibnamefont
  {Strelets}},\ }\bibfield  {title} {\enquote {\bibinfo {title} {Mechanisms of
  transition and heat transfer in a separation bubble},}\ }\href@noop {}
  {\bibfield  {journal} {\bibinfo  {journal} {J. Fluid Mech.}\ }\textbf
  {\bibinfo {volume} {403}},\ \bibinfo {pages} {329--349} (\bibinfo {year}
  {2000})}\BibitemShut {NoStop}%
\bibitem [{\citenamefont {Guezennec}, \citenamefont {Stretch},\ and\
  \citenamefont {Kim}(1990)}]{guezennec1990structure}%
  \BibitemOpen
  \bibfield  {author} {\bibinfo {author} {\bibfnamefont {Y.}~\bibnamefont
  {Guezennec}}, \bibinfo {author} {\bibfnamefont {D.}~\bibnamefont {Stretch}},
  \ and\ \bibinfo {author} {\bibfnamefont {J.}~\bibnamefont {Kim}},\ }\bibfield
   {title} {\enquote {\bibinfo {title} {The structure of turbulent channel flow
  with passive scalar transport},}\ }\href@noop {} {\bibfield  {journal}
  {\bibinfo  {journal} {Studying Turbulence Using Numerical Simulation
  Databases. 3: Proceedings of the 1990 Summer Program}\ ,\ \bibinfo {pages}
  {127--138}} (\bibinfo {year} {1990})}\BibitemShut {NoStop}%
\bibitem [{\citenamefont {Abe}\ and\ \citenamefont
  {Antonia}(2017)}]{abe2017relationship}%
  \BibitemOpen
  \bibfield  {author} {\bibinfo {author} {\bibfnamefont {H.}~\bibnamefont
  {Abe}}\ and\ \bibinfo {author} {\bibfnamefont {R.~A.}\ \bibnamefont
  {Antonia}},\ }\bibfield  {title} {\enquote {\bibinfo {title} {Relationship
  between the heat transfer law and the scalar dissipation function in a
  turbulent channel flow},}\ }\href@noop {} {\bibfield  {journal} {\bibinfo
  {journal} {J. Fluid Mech.}\ }\textbf {\bibinfo {volume} {830}},\ \bibinfo
  {pages} {300--325} (\bibinfo {year} {2017})}\BibitemShut {NoStop}%
\bibitem [{\citenamefont {Yuan}\ and\ \citenamefont
  {Piomelli}(2014)}]{yuan2014roughness}%
  \BibitemOpen
  \bibfield  {author} {\bibinfo {author} {\bibfnamefont {J.}~\bibnamefont
  {Yuan}}\ and\ \bibinfo {author} {\bibfnamefont {U.}~\bibnamefont
  {Piomelli}},\ }\bibfield  {title} {\enquote {\bibinfo {title} {Roughness
  effects on the reynolds stress budgets in near-wall turbulence},}\
  }\href@noop {} {\bibfield  {journal} {\bibinfo  {journal} {J. Fluid Mech.}\
  }\textbf {\bibinfo {volume} {760}},\ \bibinfo {pages} {R1} (\bibinfo {year}
  {2014})}\BibitemShut {NoStop}%
\bibitem [{\citenamefont {Raupach}\ and\ \citenamefont
  {Shaw}(1982)}]{raupach1982averaging}%
  \BibitemOpen
  \bibfield  {author} {\bibinfo {author} {\bibfnamefont {M.~R.}\ \bibnamefont
  {Raupach}}\ and\ \bibinfo {author} {\bibfnamefont {R.~H.}\ \bibnamefont
  {Shaw}},\ }\bibfield  {title} {\enquote {\bibinfo {title} {{Averaging
  procedures for flow within vegetation canopies}},}\ }\href@noop {} {\bibfield
   {journal} {\bibinfo  {journal} {Bound. Layer Meteorol.}\ }\textbf {\bibinfo
  {volume} {22}},\ \bibinfo {pages} {79--90} (\bibinfo {year}
  {1982})}\BibitemShut {NoStop}%
\bibitem [{\citenamefont {Ashrafian}\ and\ \citenamefont
  {Andersson}(2006)}]{ashrafian2006structure}%
  \BibitemOpen
  \bibfield  {author} {\bibinfo {author} {\bibfnamefont {A.}~\bibnamefont
  {Ashrafian}}\ and\ \bibinfo {author} {\bibfnamefont {H.~I.}\ \bibnamefont
  {Andersson}},\ }\bibfield  {title} {\enquote {\bibinfo {title} {The structure
  of turbulence in a rod-roughened channel},}\ }\href@noop {} {\bibfield
  {journal} {\bibinfo  {journal} {Int. J. Heat Fluid Flow}\ }\textbf {\bibinfo
  {volume} {27}},\ \bibinfo {pages} {65--79} (\bibinfo {year}
  {2006})}\BibitemShut {NoStop}%
\bibitem [{\citenamefont {Smalley}\ \emph {et~al.}(2002)\citenamefont
  {Smalley}, \citenamefont {Leonardi}, \citenamefont {Antonia}, \citenamefont
  {Djenidi},\ and\ \citenamefont {Orlandi}}]{smalley2002reynolds}%
  \BibitemOpen
  \bibfield  {author} {\bibinfo {author} {\bibfnamefont {R.}~\bibnamefont
  {Smalley}}, \bibinfo {author} {\bibfnamefont {S.}~\bibnamefont {Leonardi}},
  \bibinfo {author} {\bibfnamefont {R.}~\bibnamefont {Antonia}}, \bibinfo
  {author} {\bibfnamefont {L.}~\bibnamefont {Djenidi}}, \ and\ \bibinfo
  {author} {\bibfnamefont {P.}~\bibnamefont {Orlandi}},\ }\bibfield  {title}
  {\enquote {\bibinfo {title} {Reynolds stress anisotropy of turbulent rough
  wall layers},}\ }\href@noop {} {\bibfield  {journal} {\bibinfo  {journal}
  {Exp. Fluids}\ }\textbf {\bibinfo {volume} {33}},\ \bibinfo {pages} {31--37}
  (\bibinfo {year} {2002})}\BibitemShut {NoStop}%
\end{thebibliography}%
\providecommand{\noopsort}[1]{}\providecommand{\singleletter}[1]{#1}%

\end{document}